\newcommand{\be}{\begin{equation}}
\newcommand{\ee}{\end{equation}}

\documentclass{mn2e}
\input{epsf}

\title[Tidal evolution of disky dwarf galaxies]
{Tidal evolution of disky dwarf galaxies in the Milky Way potential: the formation of dwarf spheroidals}

\author[J. Klimentowski et al.]
    {Jaros{\l}aw Klimentowski,$^{1}$ Ewa L. {\L}okas,$^{1}$ Stelios Kazantzidis,$^{2}$
    Lucio Mayer$^{3,4}$ \newauthor{and Gary A. Mamon}$^{5,6}$
    \\
    \\
    $^1$Nicolaus Copernicus Astronomical Center, Bartycka 18,
    00-716 Warsaw, Poland\\
    $^2$Center for Cosmology and Astro-Particle Physics;
        and Department of Physics; and Department of Astronomy, \\
    The Ohio State University, Physics Research Building, 191 West Woodruff Avenue, Columbus, OH 43210, USA\\
    $^3$Institute for Theoretical Physics, University of Z\"urich, CH-8057 Z\"urich, Switzerland\\
    $^4$Institute of Astronomy, Department of Physics, ETH Z\"urich, Wolfgang-Pauli
    Strasse, CH-8093 Z\"urich, Switzerland  \\
    $^5$Institut d'Astrophysique de Paris (UMR 7095: CNRS and Universit\'e Pierre \& Marie Curie),
    98 bis Bd Arago, F-75014 Paris, France \\
    $^6$GEPI (UMR 8111: CNRS and Universit\'e Denis Diderot), Observatoire de Paris,
    F-92195 Meudon, France }
\begin{document}

\maketitle

\begin{abstract}
We conduct high-resolution collisionless $N$-body simulations to investigate the
tidal evolution of dwarf galaxies on an eccentric orbit in the
Milky Way (MW) potential. The dwarfs originally consist of a low surface brightness
stellar disk embedded in a cosmologically motivated dark matter halo.
During 10 Gyr of dynamical evolution and after $5$ pericentre passages the dwarfs suffer
substantial mass loss and their stellar component undergoes a major morphological
transformation from a disk to a bar and finally to a spheroid.
The bar is preserved for most of the time as the angular momentum is
transferred outside the galaxy. A dwarf spheroidal (dSph) galaxy
is formed via gradual shortening of the bar. This work thus provides a comprehensive
quantitative explanation of a potentially crucial morphological transformation mechanism for dwarf
galaxies that operates in groups as well as in clusters.
We compare three cases with different initial inclinations of the disk
and find that the evolution is fastest when the disk is coplanar with the orbit. Despite the strong tidal
perturbations and mass loss the dwarfs remain dark matter dominated.
For most of the time the 1D stellar velocity dispersion, $\sigma$, follows the maximum
circular velocity, $V_{\rm max}$, and they are both good tracers of the bound mass. Specifically, we find that
$M_{\rm bound} \propto V_{\rm max}^{3.5}$ and $V_{\rm max} \sim \sqrt{3} \sigma$ in agreement with
earlier studies based on pure dark matter simulations. The latter relation is based on directly
measuring the stellar kinematics of the simulated dwarf and may thus be reliably used to map the
observed stellar velocity dispersions of dSphs to halo circular velocities when addressing the
missing satellites problem.
\end{abstract}


\section{Introduction}

In the currently favoured cold dark matter (CDM) paradigm of hierarchical
structure formation, structure develops from the `bottom-up'
as small, dense dark matter clumps collapse first
and subsequently undergo a series of mergers that result in the hierarchical
formation of large, massive dark matter haloes. According to this model,
dwarf galaxies constitute the building blocks of larger galaxies and
those that have survived until now are expected to be among the oldest
structures in the Universe. Cosmological $N$-body simulations set within the
CDM paradigm predict too many substructures around galaxy-sized systems compared to
 the number of dwarf galaxy satellites of the Milky Way (MW) and M31 (Klypin et al. 1999; Moore et al. 1999),
giving rise to the so-called `missing satellites' problem.
It is still unclear whether this problem lies in the theory or in the inadequacy of
current observations for very faint galaxies (for a review see Kravtsov,
Gnedin \& Klypin 2004a; Simon \& Geha 2007).

Among the dwarf galaxies of the Local Group (see Mateo 1998 for a review),
dwarf spheroidals (dSph) are the most numerous. Owing to their proximity
to the primary galaxies it is entirely plausible that dSphs have been affected by tidal
interactions. Their stellar distribution is supported by velocity
dispersion and under the assumption that they are in dynamical equilibrium,
dSphs are often characterized by very high dark matter contents and
mass-to-light ratios. Though it has been suggested that such high velocity
dispersions may result from lack of virial equilibrium (Kuhn \& Miller 1989) or even
non-Newtonian dynamics like MOND (Milgrom 1995; {\L}okas 2001), the general
consensus is that dSphs are the most dark matter dominated galaxies in
the Universe (Gilmore et al. 2007).

With increasing accuracy of observational data more precise
studies of dSph dynamics can be performed.  Recently,
their nearly flat velocity dispersion profiles have been modelled
using the Jeans formalism assuming a compact stellar component embedded
in an extended dark matter halo ({\L}okas 2002; Kleyna et al. 2002).
However, there are still uncertainties concerning the exact form of the dark
matter density distribution. For example Walker et al. (2006)
exclude that dark matter follows light in the Fornax dSph, while
Klimentowski et al. (2007) argue that it is possible to fit a constant
mass-to-light ratio if unbound stars are removed from the kinematic sample.
Similar conclusions were reached by {\L}okas (2009) using much larger data samples for Fornax, Carina, Sculptor and
Sextans dwarfs.

In recent years, two types of numerical studies have been performed to
elucidate the origin and evolution of dSphs. The first is based on
cosmological $N$-body simulations that follow only the dark
matter component.  A region of the size of the Local Group or of a MW-sized
halo is selected at $z=0$, traced back to the initial
conditions and then resimulated with higher resolution (Klypin et al. 2001;
see also Kravtsov et al. 2004a; Warnick \& Knebe 2006; Diemand, Kuhlen \& Madau 2007;
Martinez-Vaquero, Yepes \& Hoffman 2007 for examples of this approach).
This strategy allows to accurately study the formation and evolution
of host haloes and their substructure. However, because the
baryonic component is neglected, it cannot address the internal structure and
kinematics of dwarfs.

The second approach is based on evolving a high-resolution numerical model
of a single dwarf galaxy placed on a representative orbit around its host.
In most cases, the host galaxy is represented by a static external potential
(e.g. Piatek \& Pryor 1995; Johnston, Sigurdsson \& Hernquist 1999; Mayer et al. 2001; Hayashi et
al. 2003; Kazantzidis, Magorrian \& Moore 2004a; Kazantzidis et al. 2004b; Helmi 2004;
Klimentowski et al. 2007; Pe\~narrubia, Navarro \& McConnachie 2008).
With this method one readily isolates the effects of tidal interactions
on the internal structure and kinematics of the dwarf. However, this
approach requires making assumptions regarding the initial conditions
and neglects the effects of dynamical friction, the evolution of the host
galaxy's potential, or interactions between dwarfs.

\begin{figure}
    \leavevmode
    \epsfxsize=7.2cm
    \epsfbox[0 0 330 370]{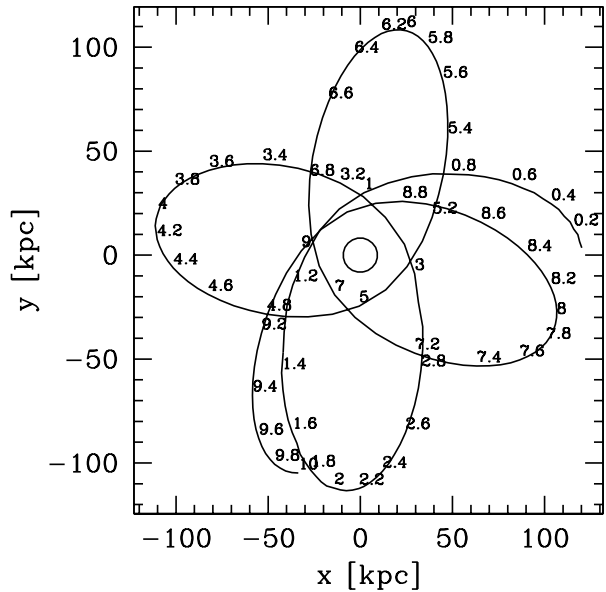}
\caption{The orbit of the simulated dwarf galaxy projected on the plane of
the MW disk. The simulation starts at the right-hand side of the Figure. Numbers indicate
the time from the start of the simulation in Gyr.
The circle in the middle shows the position of the host
galaxy. }
\label{orbit}
\end{figure}

In this paper we adopt the second approach. Our initial conditions are based
on the `tidal stirring' model (Mayer et al. 2001). In particular, the
progenitor dwarfs comprise a low surface brightness stellar disk embedded
in a dark matter halo, and they are placed on an eccentric orbit within a static
potential representing the MW. We aim at elucidating the dynamical
mechanisms whereby disky dwarfs can be transformed into spheroidal
galaxies and how the evolution occurs. In addition, we investigate the relation
between the stellar velocity dispersion, the maximum circular velocity and the
bound mass of the dwarfs as a function of time. Knowing these relations is
crucial for interpreting correctly the missing satellites problem as they
reflect the link between dark halo properties and observable
properties of dSphs.

The paper is organized as follows. In section 2 we summarize the simulation details.
In section 3 we study an example of galaxy evolution in terms of the mass loss, morphological
transformation and changes in internal kinematics. Section 4 discusses the dependence
of the results on the initial orientation of the dwarf galaxy disk and section 5 is devoted
to observational consequences of the tidal stirring scenario. The discussion follows in section 6.

\begin{figure*}
    \leavevmode
    \epsfxsize=16cm
    \epsfbox{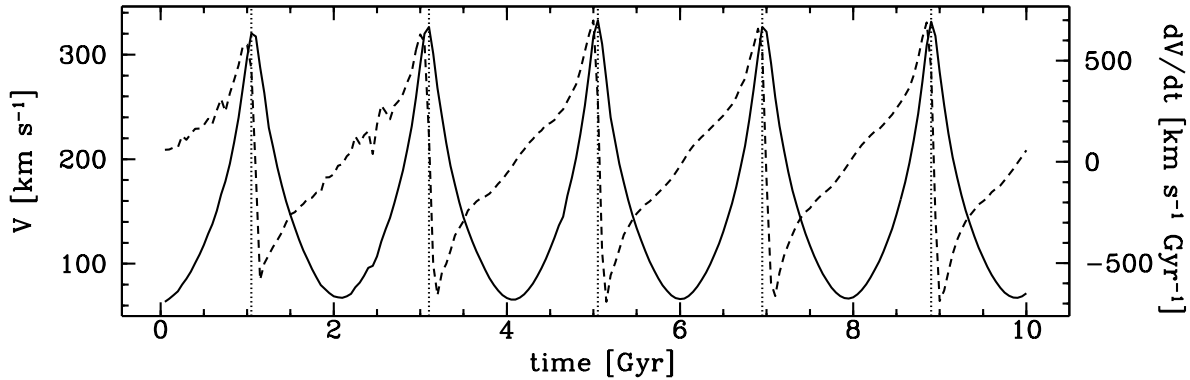}
\caption{The modulus of the velocity vector of the dwarf galaxy (solid line, left axis)
	and its acceleration (dashed line, right axis) as a function of time. Vertical
	dotted lines indicate pericentre passages.}
\label{velocity}
\end{figure*}

\section{The simulations}

In this section we summarize the most important features of the simulations used for this
study (see Klimentowski et al. 2007 for details). The simulations were designed
according to the `tidal stirring' scenario proposed by Mayer et al. (2001) which suggests that
the progenitors of the majority of dSph galaxies were rotationally supported
low surface brightness disky dwarfs similar to present-day dwarf irregulars.

The dwarf progenitors consisted of a stellar disk embedded in a dark matter halo.
The disk is modelled by $N = 10^6$ particles of mass $149.5$ M$_{\odot}$.
Its density drops exponentially with radius in the
rotation plane and its vertical structure is modelled by isothermal sheets. The dark matter halo consists
of $N = 4 \times 10^6$ particles of mass $1035.7$ M$_{\odot}$ which are distributed according to the
NFW profile. The concentration parameter of the NFW halo is $c=15$.
Beyond the virial radius the density profile was truncated with an exponential function
to keep the total mass finite (Kazantzidis et al. 2004a). The total mass of
the progenitor was $M = 4 \times 10^9$ M$_{\odot}$ (the virial mass is $M_{\rm vir} = 3.7
\times 10^9$ M$_{\odot}$) which corresponds to
the virial velocity of 20 km s$^{-1}$. The models include adiabatic contraction of the halo in response
to the baryons (Blumenthal et al. 1986). The peak velocity $V_{\rm max}$ is initially equal to 30 km s$^{-1}$.

The dwarf galaxies evolve on their orbit around the host galaxy
which is modelled by a static gravitational potential. This approach neglects
several effects like dynamical friction or evolution
of the host galaxy itself. However, for our setup the timescale of the dynamical friction should
significantly exceed the orbital timescale (Colpi, Mayer \& Governato 1999), while the host halo should be
already in place by the time the dwarf enters it. Indeed cosmological hydrodynamical
simulations that model the formation of the MW suggest that it had already accreted most of its
dark matter and baryonic mass between $z=1$ and $z=2$, i.e. between 8 and 10 Gyr ago (Governato
et al. 2007). For simplicity, the potential of the host galaxy is thus assumed
to have the present-day properties of the MW. We use a static NFW halo
with a virial mass of $M_{\rm vir} = 10^{12}$ M$_{\odot}$. We also add the potential of
a stellar disk with mass $M_{\rm D} = 4 \times 10^{10}$ M$_{\odot}$.

We choose a single, typical cosmological orbit with apocentre to pericentre ratio of $r_{\rm a}/r_{\rm p} \approx 5$
(e.g. Ghigna et al. 1998).
With a fairly small pericentre of 22 kpc the orbit should be quite typical among those
of subhaloes falling into the MW halo quite early, around $z=2$ (Diemand et al. 2007;
Mayer et al. 2007). Our default simulation (described in Klimentowski et al. 2007) had the disk of the
dwarf galaxy initially oriented perpendicular to the orbital plane. For the purpose of the present
study we ran two additional simulations with the disk lying in the orbital plane and at the inclination of 45
deg (and all the other simulation parameters, including the orbit, unchanged).
In both new cases the motion of stars in the disk was prograde with respect to the orbital motion of
the dwarf.

The simulations were performed using PKDGRAV, a multi-stepping, parallel, tree $N$-body  code
(Stadel 2001). The gravitational softening length was 50 pc for stars and 100 pc for dark matter.
We used 200 outputs of each simulation saved at equal snapshots of 0.05 Gyr.


\section{Galaxy evolution}

\subsection{The orbit}

In this section we discuss the results of our default or reference simulation where the disk of the dwarf
was initially oriented perpendicular to the orbital plane. The dependence on the initial orientation of the
disk is discussed in section 4.

\begin{figure*}
    \leavevmode
    \epsfxsize=14cm
    \epsfbox[20 0 518 183]{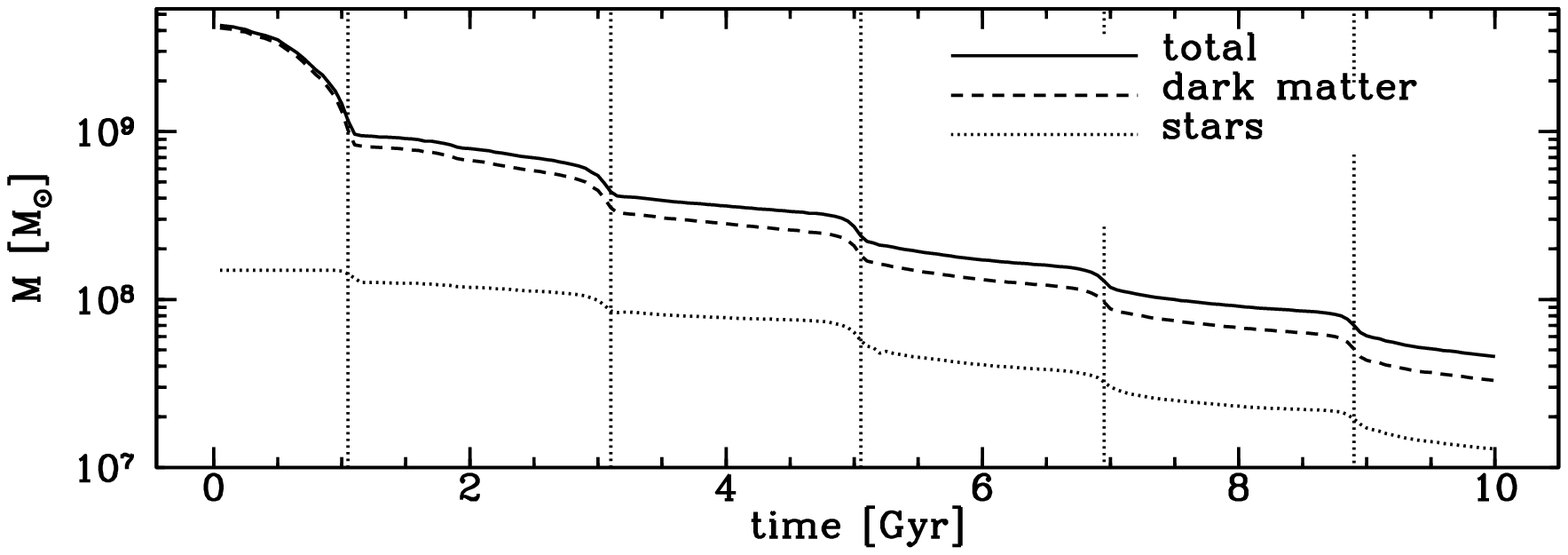}
    \leavevmode
    \epsfxsize=14cm
    \epsfbox[20 0 518 183]{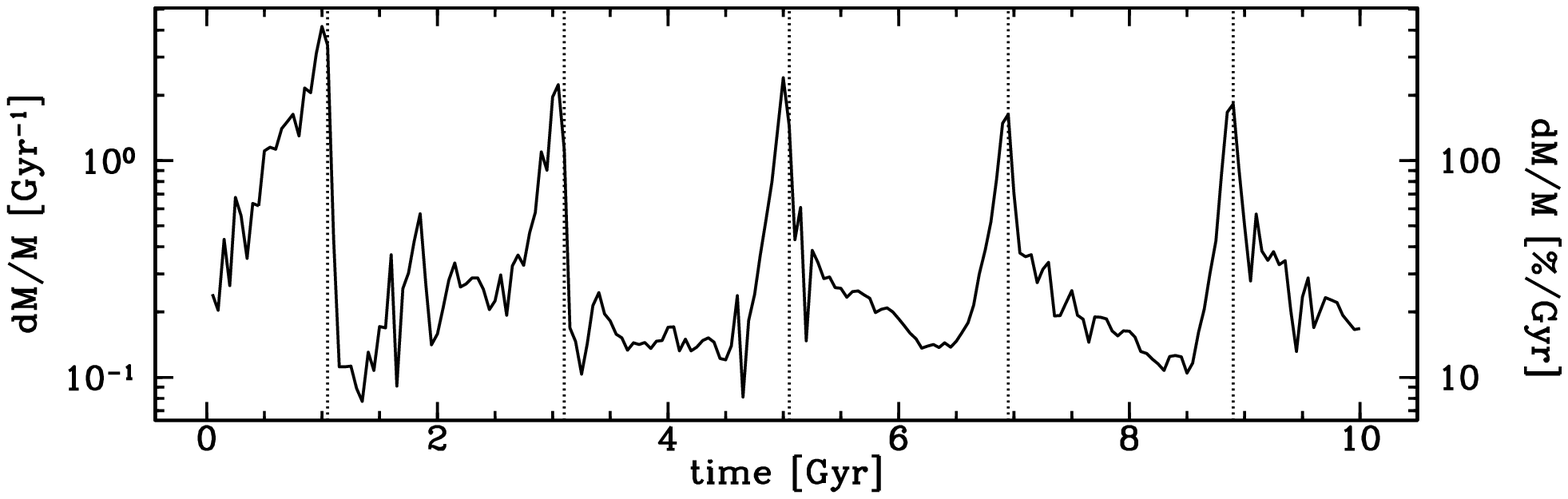}
    \leavevmode
    \epsfxsize=14cm
    \epsfbox[20 0 518 183]{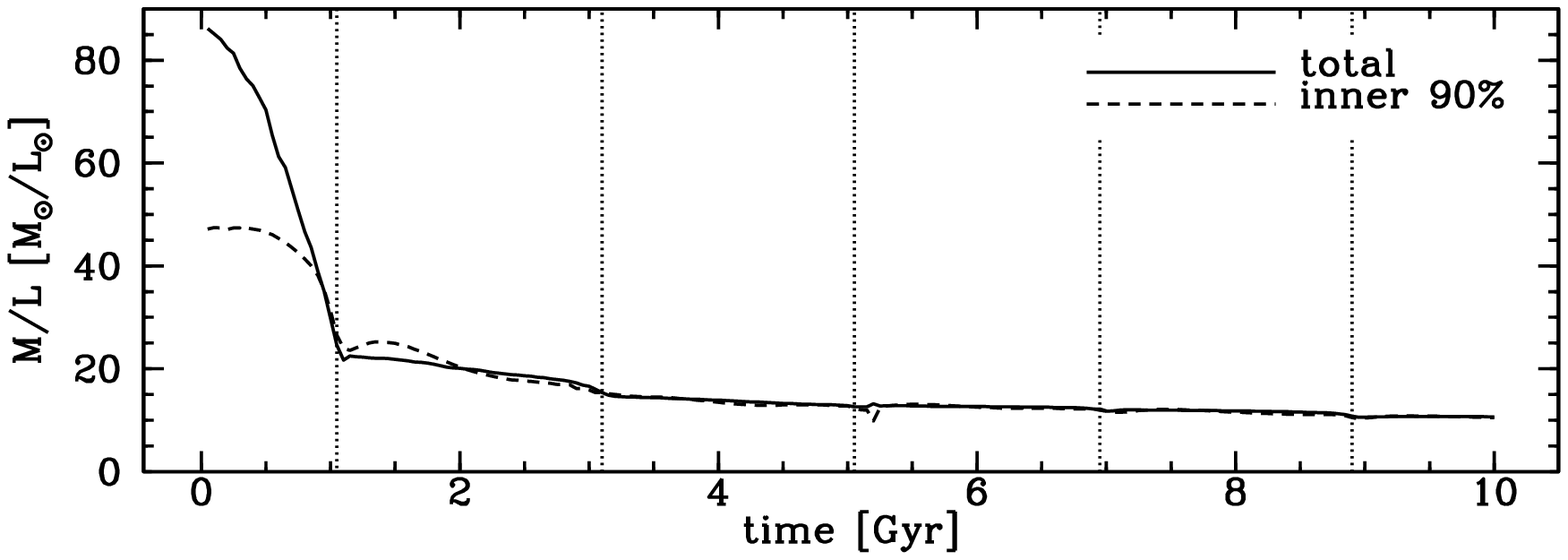}
\caption{The evolution of the mass and mass-to-light ratio of the dwarf galaxy in time.
	In the upper panel we plot the mass of all bound particles (solid line), dark matter
	(dashed line) and stars (dotted line). The middle panel shows the relative mass loss.
	The lower panel plots the
	mass-to-light ratio assuming the stellar mass-to-light ratio of 3 solar units. The solid
	line presents the results for all bound particles, the dashed line plots the same quantity measured inside
	a sphere containing 90 percent of bound stars. In all panels vertical dotted lines
	indicate pericentre passages.}
\label{massloss}
\end{figure*}

Fig.~\ref{orbit} shows the orbit of the simulated dwarf
galaxy during its whole evolution projected on the orbital $xy$ plane, the
plane of the stellar disk of the host galaxy.
The simulation started at the right-hand side of the Figure, at the apocentre, 120 kpc away from the host galaxy.
Numbers correspond to the evolution time in Gyr since the start of the simulation.
The simulation ends after 10 Gyr of evolution at the bottom of the Figure. There are 5 pericentres
during the evolution. Due to the lack of dynamical friction all orbits have very similar
parameters, except for the in-plane precession which amounts to about 90 deg
per orbit. Small changes are due to the fact that as the dwarf is stripped the matter around it
induces small fluctuations on its orbital dynamics.
The pericentre distance $r_{\rm p}$ varies from 21.7 kpc to 23.7 kpc, while the apocentre $r_{\rm a}$
from 110.9 kpc to 113.6 kpc. The ratio remains $r_{\rm a}/r_{\rm p} \approx 5$
as for typical cosmological haloes (e.g. Ghigna et al. 1998).
Figure~\ref{velocity} shows the total velocity of the galaxy and its acceleration. All
orbit passages are very similar in terms of these parameters, as expected based on the absence
of dynamical friction.

\subsection{Mass loss}

In order to transform a stellar disk into a spheroidal component the galaxy
needs to be strongly affected by tides, and several pericentre passages are required in order
for that to happen (Mayer et al. 2001, 2007). Prolonged mass loss is
unavoidable since each subsequent tidal shock weakens
the potential well of the dwarf. The amount of matter lost
depends strongly on the orbital parameters and the structural properties of
the dwarf galaxy (e.g. Hayashi et al. 2003; Kazantzidis et al. 2004a,b;
Pe\~narrubia, McConnachie \& Navarro 2007). As consecutive orbits in the
simulation are very similar, the dwarf is affected by a similar tidal field in
each orbit. Thus, we can study different stages of the evolution of the dwarf
galaxy without the complication of an orbit decaying because of dynamical friction.

In order to calculate the bound mass we need to define the bound particle. For simplicity
we decided to treat the dwarf galaxy as an
isolated object and define unbound particles as the ones with velocity greater than
the escape velocity from the gravitational potential of the dwarf galaxy. In order to speed up the
calculations we used the treecode to estimate gravitational potentials for all
particles in all snapshots.

The upper panel of Fig.~\ref{massloss} shows the mass of the galaxy
as a function of time, while the middle panel of this Figure shows the relative change of
the total mass. Mass loss increases dramatically at the pericentres due to
tidal shocking. For example, during the
second pericentre passage in a very short time three times more matter is lost than during
the rest of the whole second orbit. The first
pericentre passage is even more devastating, stripping more than two-thirds of initial galaxy mass.
Eventually, after 10 Gyr of evolution 99 percent of the initial mass is lost from the dwarf.
This is consistent with analytical models of the initial tidal truncation and subsequent
tidal shocks (Taylor \& Babul 2001; Taffoni et al. 2003).

The lower panel of Fig.~\ref{massloss} plots the mass-to-light ratio calculated assuming the stellar mass-to-light
of 3 M$_{\sun}$/L$_{\sun}$. This should be true for the present time, but not necessarily for the whole
evolution since the dwarf might undergo periodic bursts of star formation (Mayer et al. 2001). However,
for dwarfs falling in early, 10 Gyr ago as assumed here, most of the gas will be quickly stripped by
ram pressure aided by the cosmic ionizing background radiation (Mayer et al. 2007), so that star
formation will cease soon after the dwarf approaches the MW.
We present the values obtained by dividing the total bound mass by the total light as well as the
values measured inside the radius containing 90 percent of the bound stars. We can see that in both cases
the ratio decreases
rapidly in the early stages, as the extended dark matter halo is disrupted easily while not many stars are
lost. Note however that
during the subsequent evolution the stellar mass loss traces the mass loss of dark matter. Starting from
the second pericentre passage the mass-to-light ratio decreases slowly and relaxes at the value of about 10, remaining
constant till the end of the simulation.

\subsection{The circular velocity}

\begin{figure}
    \leavevmode
    \epsfxsize=7.2cm
    \epsfbox[0 0 330 370]{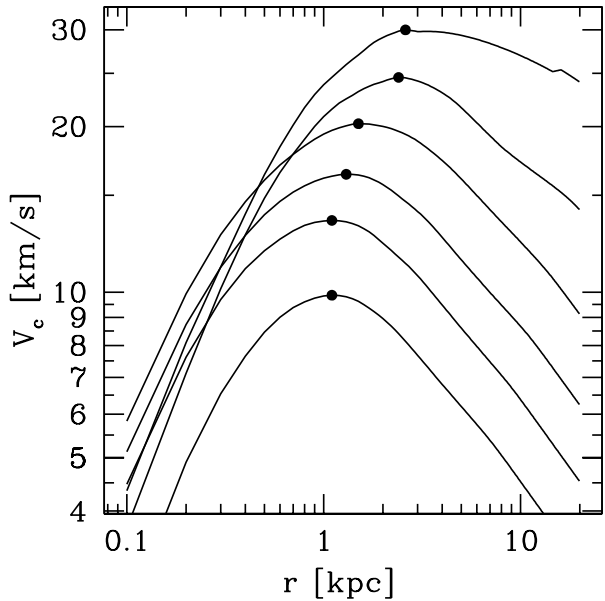}
\caption{Circular velocity profiles of the dwarf galaxy at apocentres. The highest curve is for the first
apocentre (beginning of the simulation), the lowest for the last (end of the simulation). Dots indicate the maximum
values $V_{\rm max}$.}
\label{vc}
\end{figure}

\begin{figure*}
    \leavevmode
    \epsfxsize=14cm
    \epsfbox[20 0 518 183]{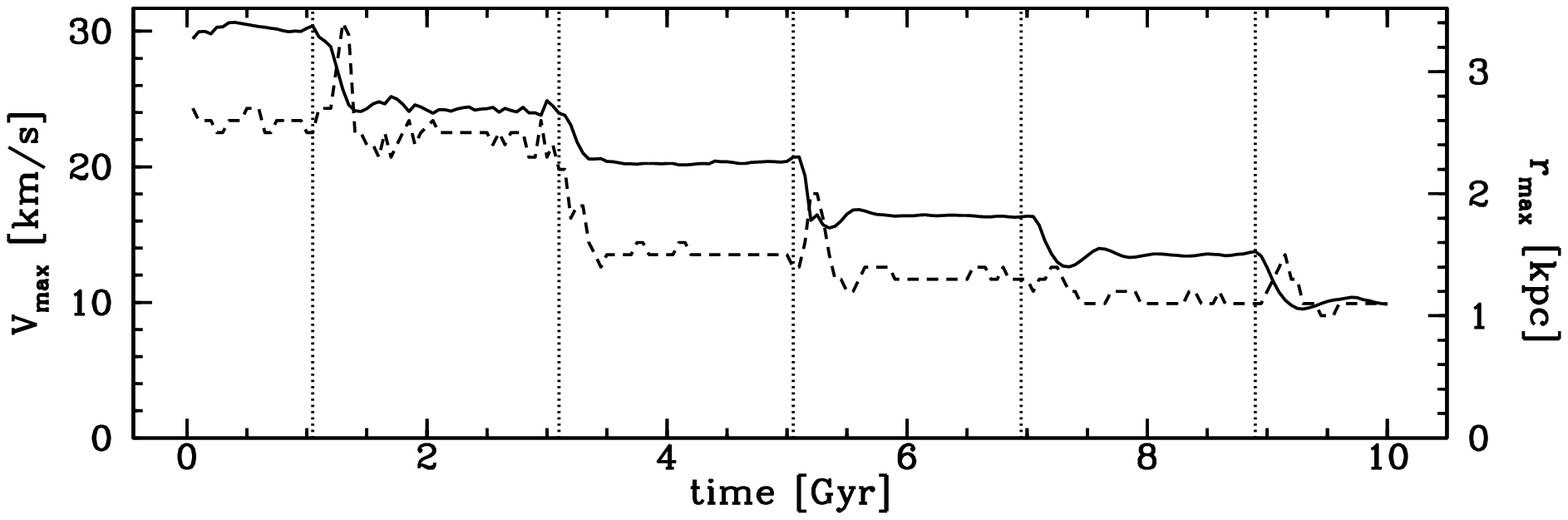}
    \epsfxsize=14cm
    \epsfbox[20 0 518 183]{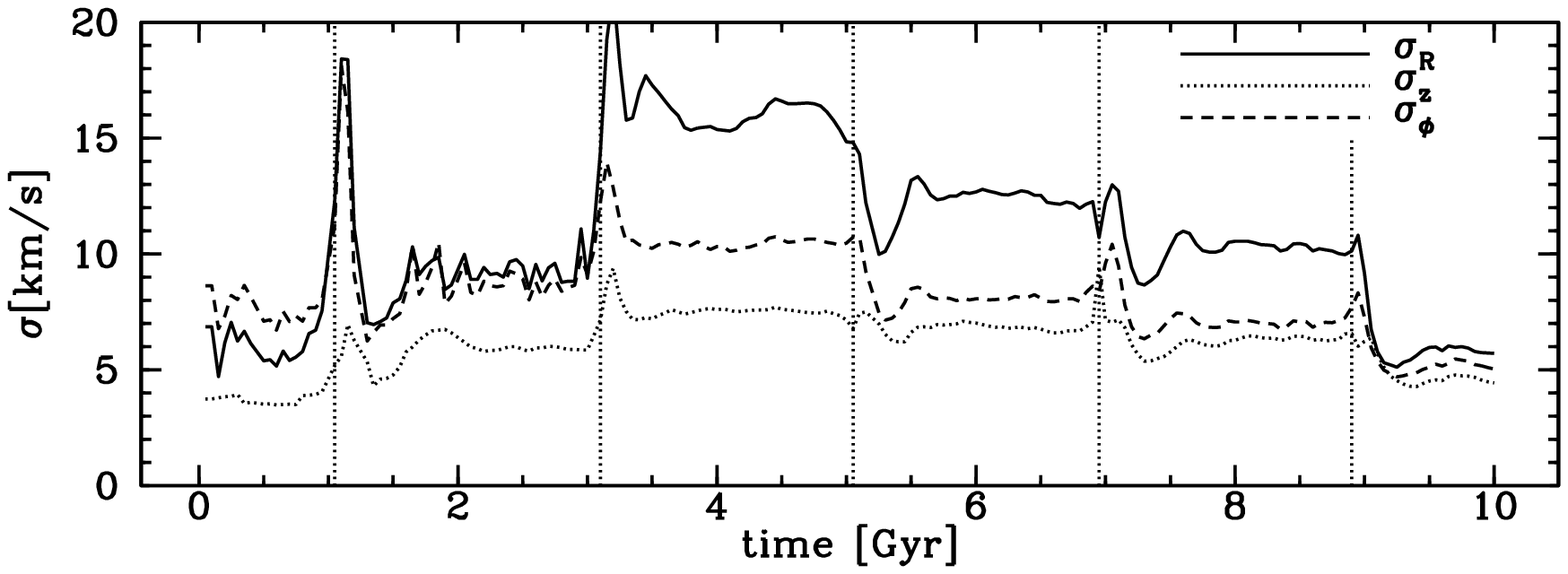}
    \epsfxsize=14cm
    \epsfbox[17 0 544 183]{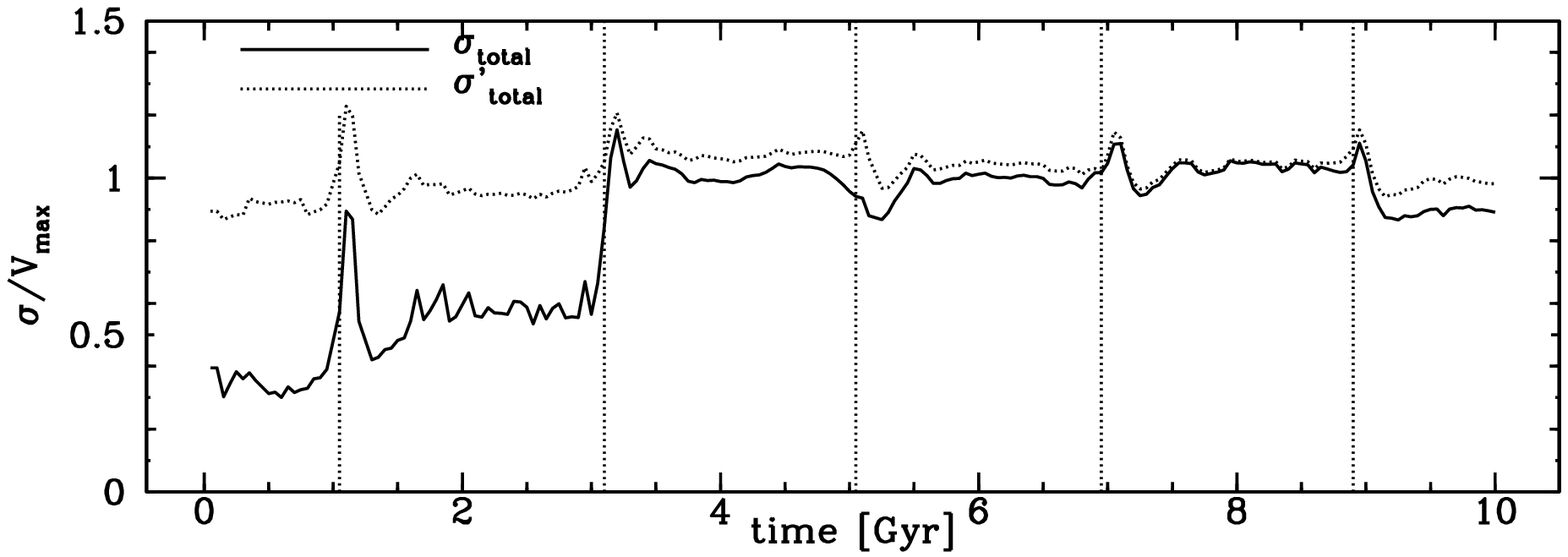}
    \caption{Upper panel: the maximum circular velocity (solid line, left axis) and the radius
	at which the maximum circular velocity occurs (dashed line, right axis). Middle panel:
	velocity dispersions of the stars within radius $r_{\rm max}$ corresponding to $V_{\rm max}$
	measured in cylindrical coordinates. Lower panel:
	the total velocity dispersion $\sigma_{\rm total}$ and $\sigma'_{\rm total}$ expressed in units of
	the maximum circular velocity.
	In all panels vertical dotted lines indicate pericentre passages.}
    \label{veldisp}
\end{figure*}

Although the bound mass is a real physical parameter it is more convenient to express the mass in terms
of the maximum circular velocity of the galaxy $V_{\rm max}$, which, as we show below, is also closely related
to the stellar velocity dispersion. This is the standard method used in cosmological simulations (e.g.
Moore et al. 1999; Stoehr et al. 2002; Kravtsov et al. 2004a; Diemand et al. 2007). It is based on the assumption that
$V_{\rm max}$ changes less than other subhalo parameters during its evolution, making it a good choice
for tracking subhaloes themselves. Figure~\ref{vc} shows the profiles of the circular velocity
$V_{\rm c}=[G M(r)/r]^{1/2}$ of the dwarf at the six apocentres with dots marking the maximum values $V_{\rm max}$.
The upper panel of Fig.~\ref{veldisp} shows the evolution of $V_{\rm max}$ in time
and the radius $r_{\rm max}$ at which this velocity occurs.

\begin{figure*}
    \leavevmode
    \epsfxsize=7.2cm
    \epsfbox{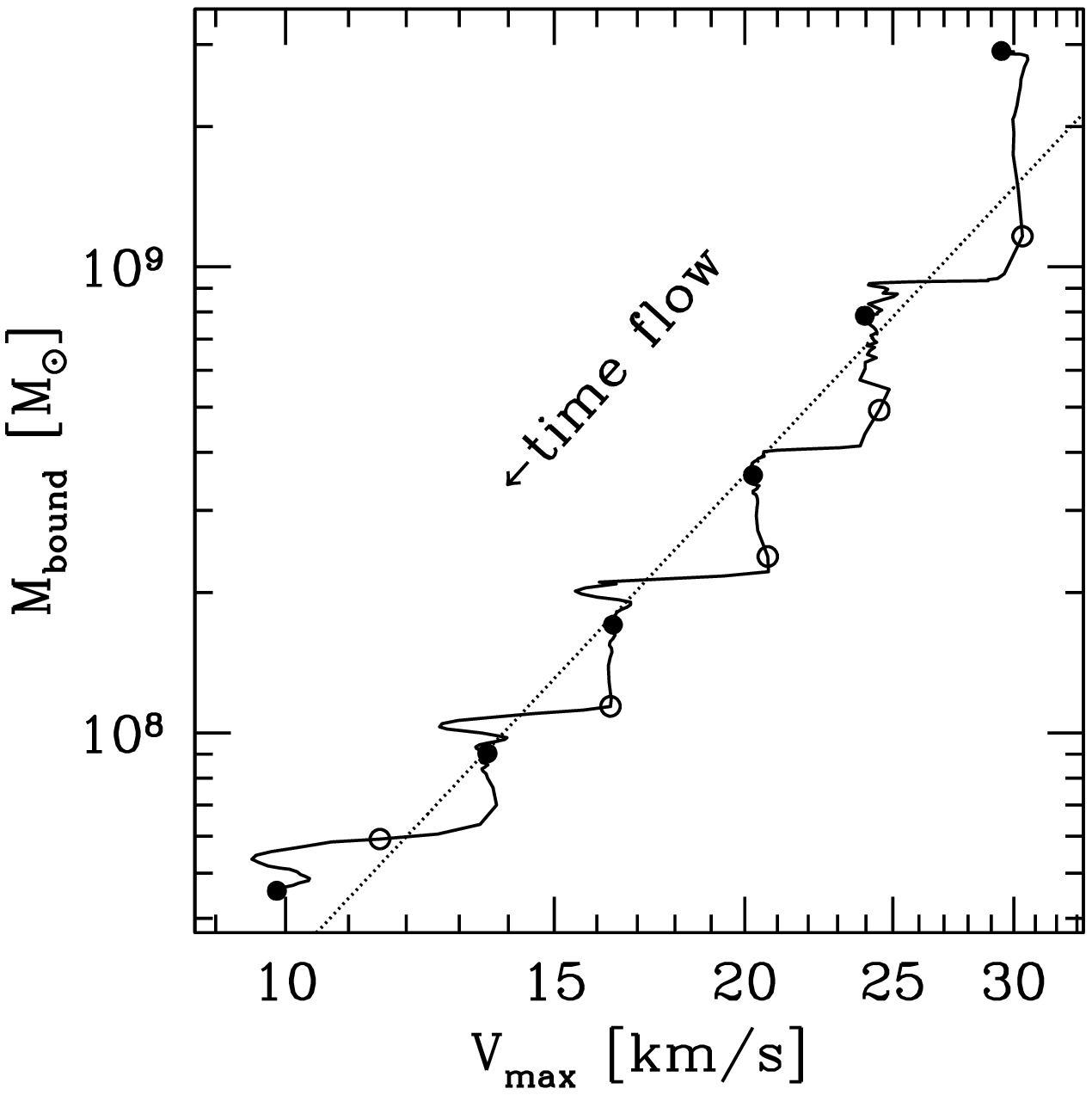}
    \epsfxsize=7.2cm
    \epsfbox{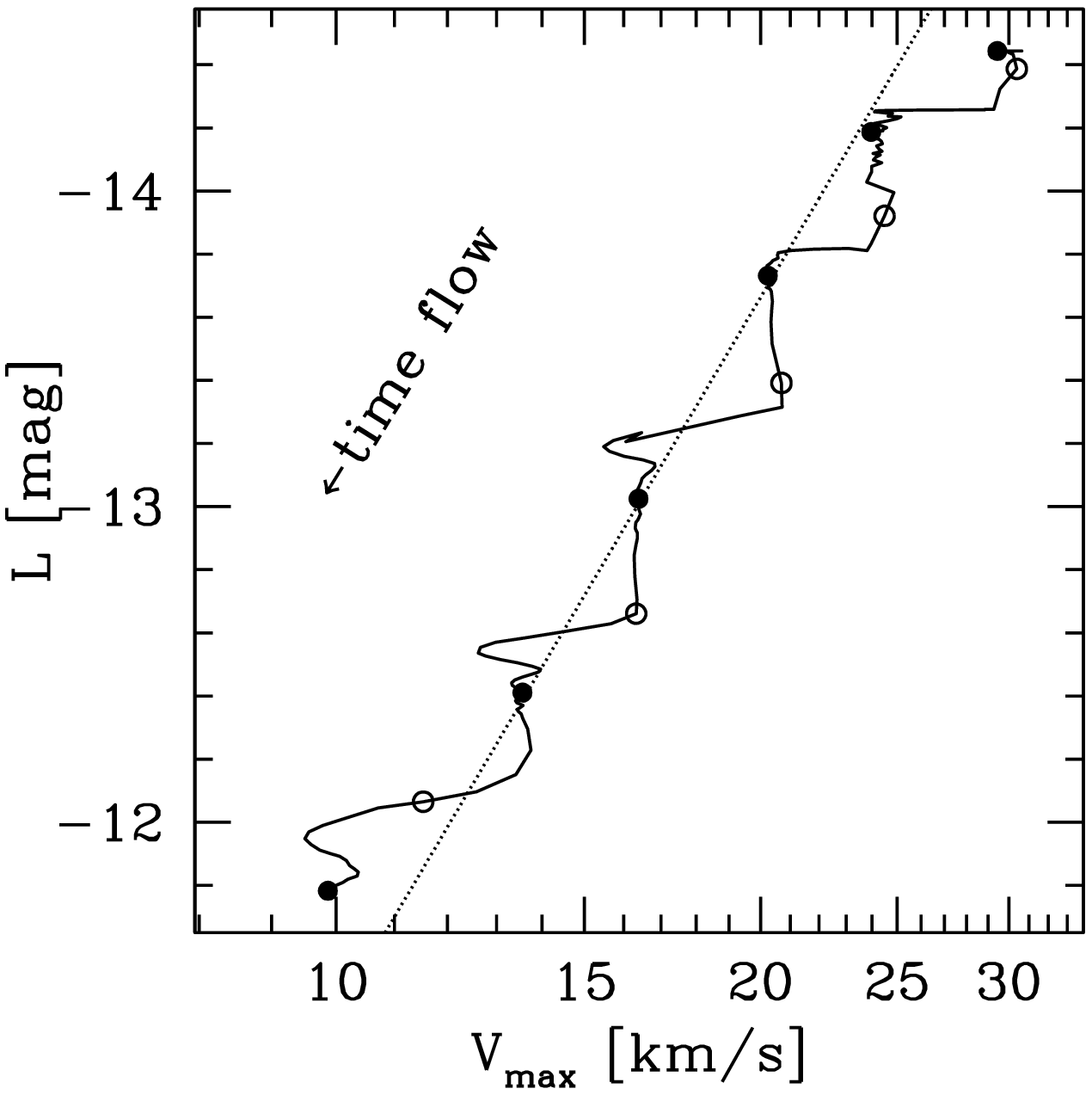}
    \caption{Mass-maximum circular velocity relations. The left panel shows the total bound mass versus the
    maximum circular velocity for the whole evolutionary
    path of the dwarf galaxy (solid line). The dotted line corresponds to the relation
    $M_{\rm bound} \propto V_{\rm max}^{3.5}$.
    The right panel shows the mass of the stellar component expressed in terms of the
    bolometric magnitude (assuming the stellar mass-to-light ratio of 3 solar units) as a function of
    the maximum circular velocity, with
    dotted line corresponding to the relation $L \propto V_{\rm max}^3$. In both panels apocentres are
    marked as filled circles and pericentres as open circles.}
    \label{velmaslum}
\end{figure*}

It is interesting to note the main difference between $V_{\rm max}$ and the total bound mass.
While the bound mass decreases during the whole
evolution, $V_{\rm max}$ remains rather constant between the pericentre passages, decreasing only shortly after the
passage, at the time when the tidal tails are formed. This suggests that $V_{\rm max}$ may be a better
measure of the mass as it is not affected by the formation of tidal tails farther out. On the other end,
$V_{\rm max}$ decreases by almost a factor of $3$ during the orbital evolution, similar to the
case of the most heavily stripped dark matter subhaloes in Kazantzidis et
al. (2004b) and Kravtsov et al. (2004a). This happens despite the
presence of the baryons which tend to moderate the effect of tidal shocks by
making the potential well deeper, especially once bar formation occurs (see
below and Mayer et al. 2007). Hence, for subhaloes that have fallen in early and
completed several orbits with fairly small pericentre passages, $V_{\rm max}$
is expected to evolve significantly. Therefore although haloes of
present-day dSphs are consistent with having $V_{\rm max}$ in
the range $15-30$ km s$^{-1}$ (Kazantzidis et al. 2004b; Strigari et al. 2007) they
could be much more massive in the past when they entered the MW halo.
Such massive dwarfs were likely not affected by reionization,
contrary to some suggestions (e.g. Grebel, Gallagher \& Harbeck 2003).

The left panel of Fig.~\ref{velmaslum} shows the
relation between $V_{\rm max}$ and the total bound mass $M_{\rm bound}$
during the entire course of dynamical evolution. The solid line presents
the whole evolutionary path and the dotted one the relation
$M_{\rm bound} \propto V_{\rm max}^\alpha$, where $\alpha=3.5$. Remarkably, although
during the 10 Gyr of evolution the dwarf galaxy loses $\sim 99$ percent of its mass
and its $V_{\rm max}$ decreases by a factor of $3$, the galaxy moves roughly
along the $M_{\rm bound} \propto V_{\rm max}^{3.5}$ line. As expected, the
largest (smallest) deviations from this relation occur at orbital
pericentres (apocentres). Our findings are in agreement with
Hayashi et al. (2003) who found that $M_{\rm bound} \propto V_{\rm max}^3$
in numerical simulations of tidal stripping of {\it pure} NFW subhaloes and
Kravtsov et al. (2004a) who reported a value of $\alpha=3.3$ for stripped
dark matter subhaloes in cosmological $N$-body simulations.

A similar comparison can be made in terms of the stellar mass instead
of the total bound mass. The right panel of Fig.~\ref{velmaslum} shows the relation
between the total luminosity of the stars and $V_{\rm max}$,
$L \propto V_{\rm max}^\alpha$. The luminosity was calculated using the bound
stellar mass and assuming a mass-to-light ratio of the stellar component
of 3 M$_{\odot}/$L$_{\odot}$ as before. This time the power-law relation
is followed only in the intermediate stages with $\alpha=3$, while
it flattens in the beginning and at the end of the simulation. A similar comparison was done
by Kravtsov et al. (2004b) for dark matter haloes in their $N$-body simulations. They
used SDSS luminosity function to predict stellar luminosities in the $R$ band for
their haloes. In the range 100 km s$^{-1} < V_{\rm max} < 200$ km s$^{-1}$
their relation also follows a power-law with $\alpha=3$. It would be difficult
to compare exact fits of the relation as one would need to apply
proper bolometric corrections, which differ for different types of stars.

\subsection{Internal kinematics and shape}

For all kinematic
measurements instead of using all bound stars we restricted the analysis to stars within
the maximum radii $r_{\rm max}$ corresponding to the maximum circular velocity $V_{\rm max}$. In the final output
this value corresponds to the radius $r=1.1$ kpc which is well within the main stellar body of the galaxy. In this
way we avoid the contamination by stars from tidal tails which may still be bound to the dwarf but are no longer
good tracers of the potential.

The middle panel of Fig.~\ref{veldisp} shows the evolution of the velocity dispersions of the
stellar particles measured
in cylindrical coordinates $R$, $z$ and $\phi$. The $z$ axis is always along the total
angular momentum vector of the stars thus the $R$-$\phi$ is the rotation plane. All velocities were
projected on a given axis and their mean values were computed (so for example for the $\phi$ direction
the mean is the average rotation velocity $V_{\phi}$).
Then the dispersions were calculated for all stars within the adopted maximum radius
with respect to the corresponding
mean values. It is interesting to note that all three velocity dispersions relax very
quickly after each pericentre and remain roughly constant for the rest of the orbit.
This suggests that the galaxy relaxes very quickly to a new equilibrium
configuration and can thus be successfully modelled using the virial theorem
or the Jeans equations for most of the time (Klimentowski et al. 2007).

\begin{figure*}
    \leavevmode
    \epsfxsize=16cm
    \epsfbox{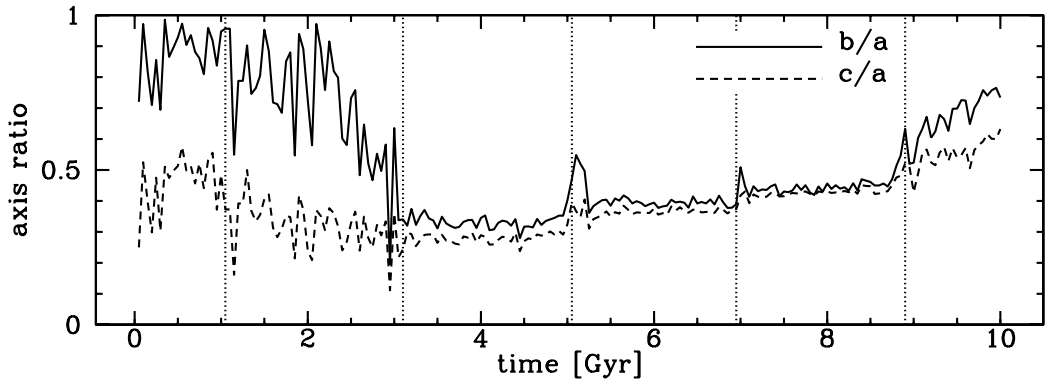}
\caption{The evolution of the shape of the stellar component of the dwarf in time.
	The shape is described by the ratios of the lengths of principal axes measured
	at the radius of maximum circular velocity (with $a,b,c$ being the longest, middle and shortest
	axis respectively).
	The vertical dotted lines indicate pericentre passages. }
\label{shape}
\end{figure*}

\begin{figure*}
\begin{center}
    \leavevmode
    \epsfxsize=14.8cm
    \epsfbox[100 40 523 170]{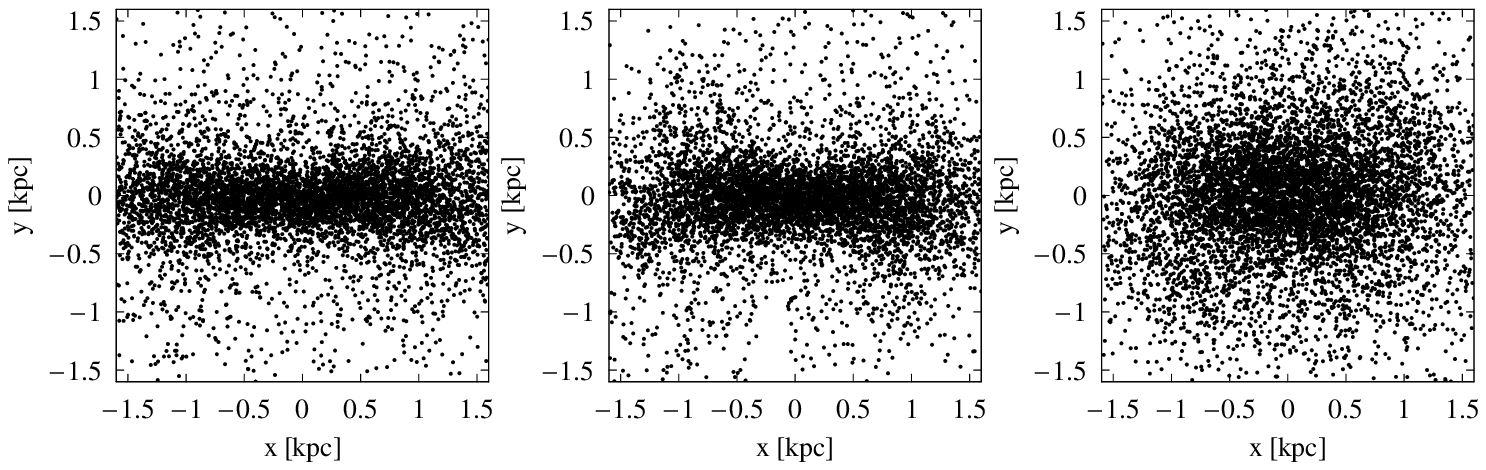}
    \epsfxsize=14.8cm
    \epsfbox[0 10 480 170]{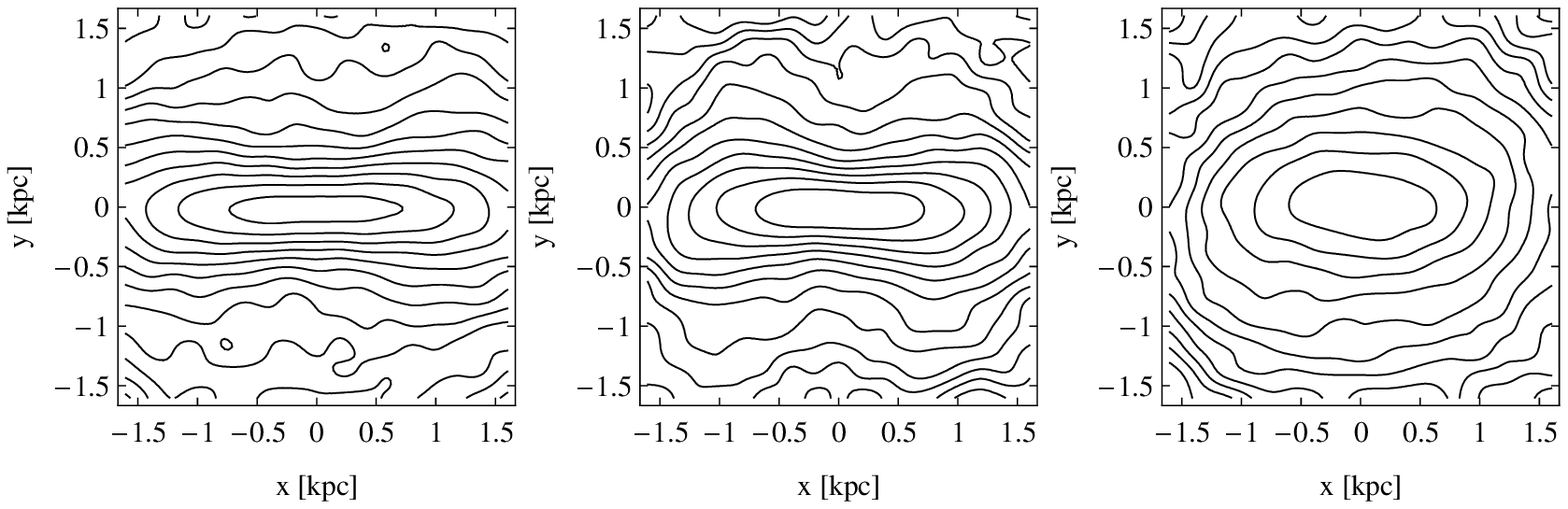}
\end{center}
\caption{Surface density distribution of the stars at 6, 8 and 10 Gyr from the beginning of the
simulation (from the left to the right column). Upper panels show the distribution of $\sim$7000
stars corresponding to 3, 5 and 10 percent of all stellar particles within $r=2.5$ kpc from the
centre of the dwarf. Lower panels plot the contours of the surface distribution $\Sigma$ expressed
in M$_{\odot}$ kpc$^{-2}$. The contours are spaced by $\Delta\log \Sigma =0.2$. The innermost
contour level is $\log \Sigma =7.2$ (lower left panel), 7 (middle) and 6.6 (right). }
\label{surface}
\end{figure*}

\begin{figure*}
\begin{center}
\begin{tabular}{lll}
    \leavevmode
    \epsfxsize=4.6cm
    \epsfbox{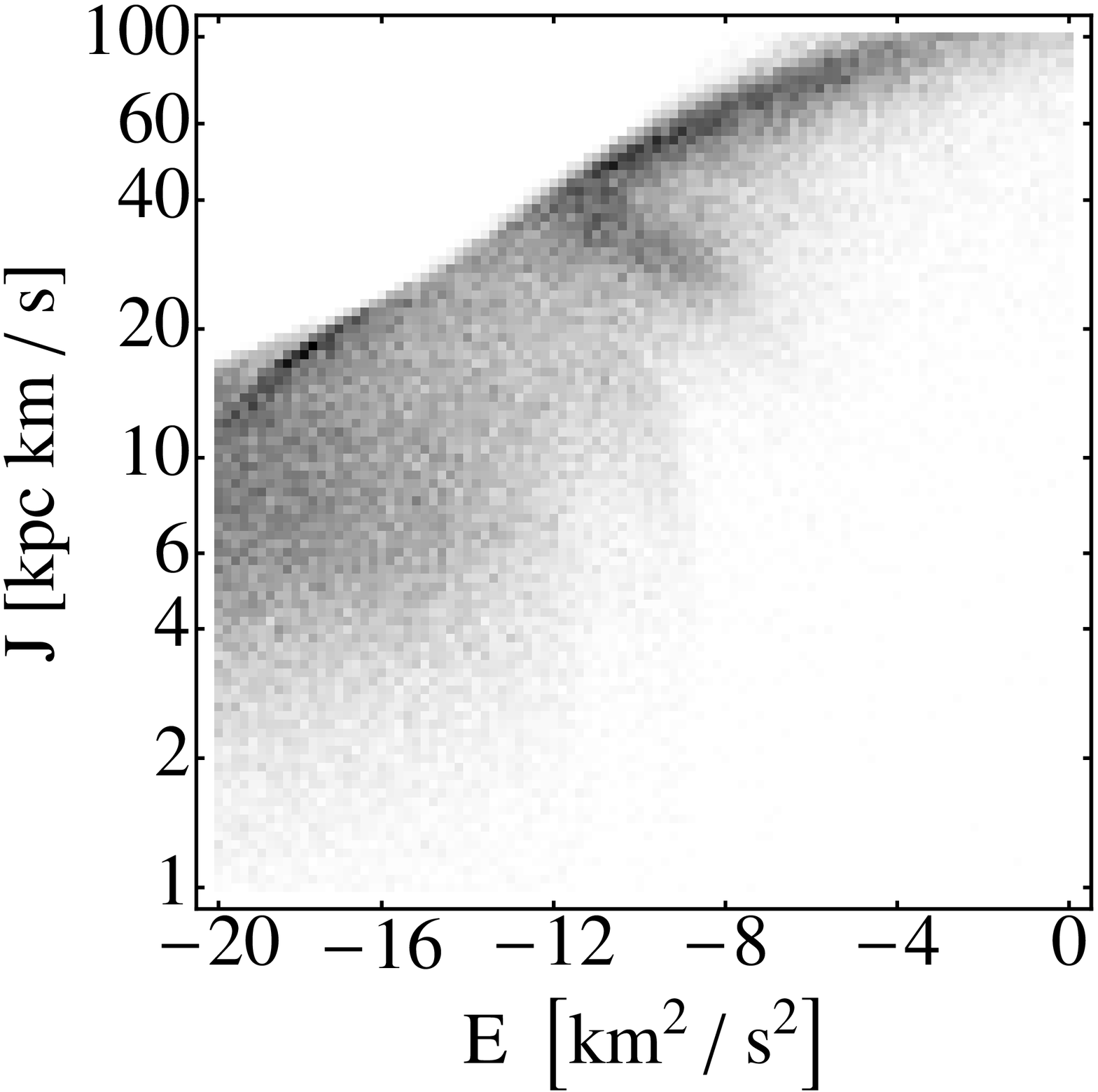} &
    \epsfxsize=4.6cm
    \epsfbox{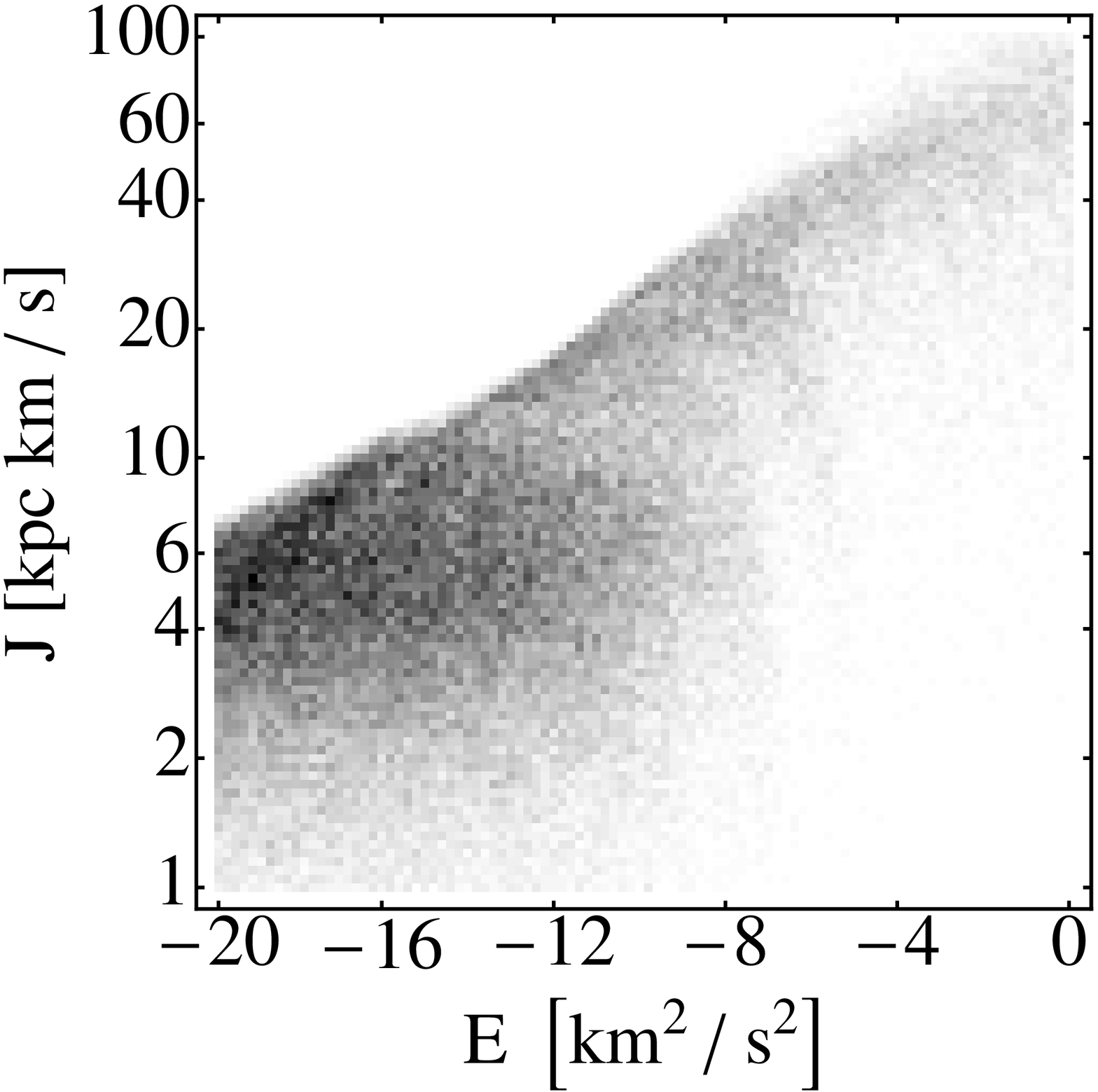} &
    \epsfxsize=4.6cm
    \epsfbox{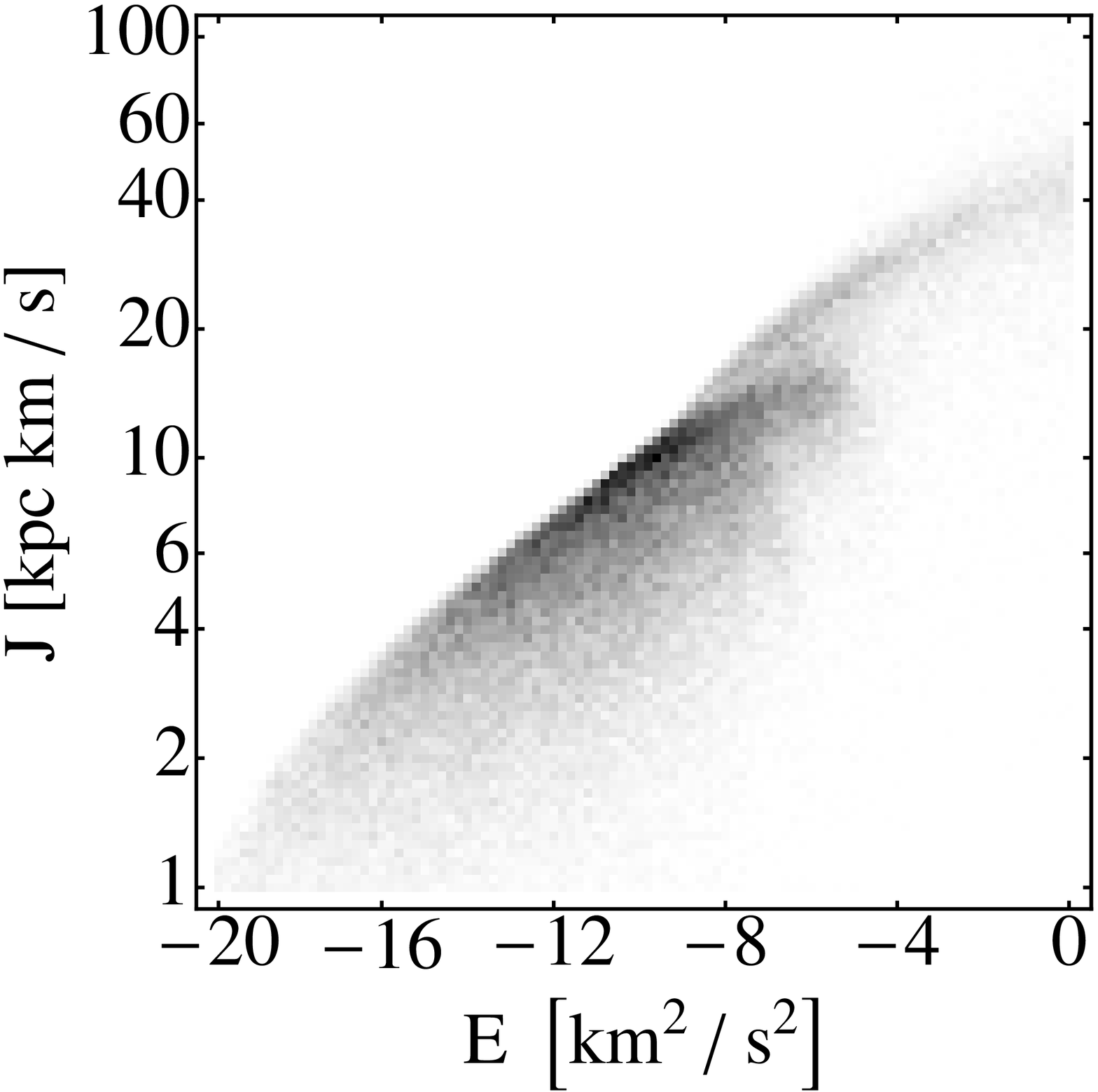} \\
    \leavevmode
    \epsfxsize=4.6cm
    \epsfbox{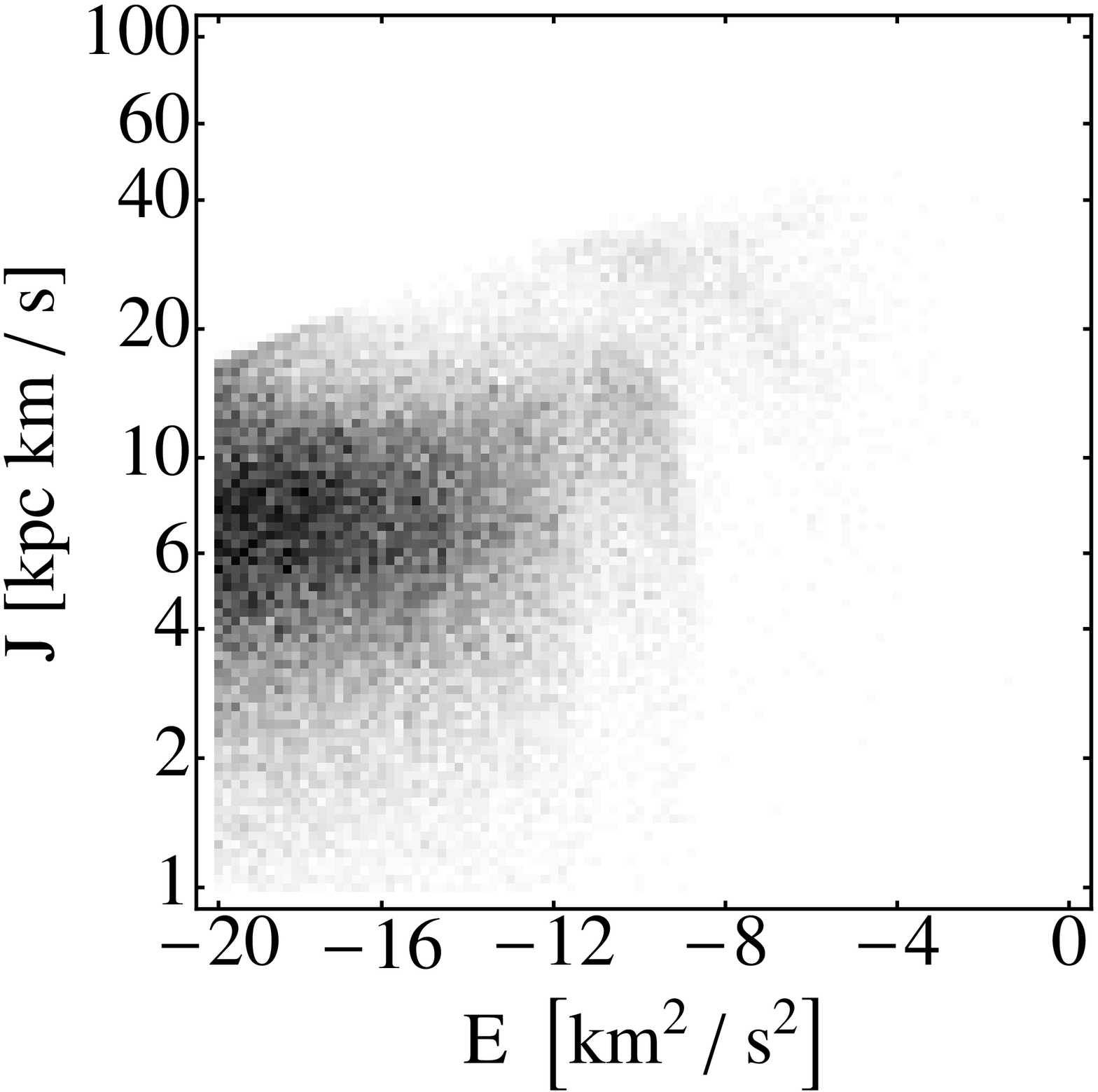}    &
    \epsfxsize=4.6cm
    \epsfbox{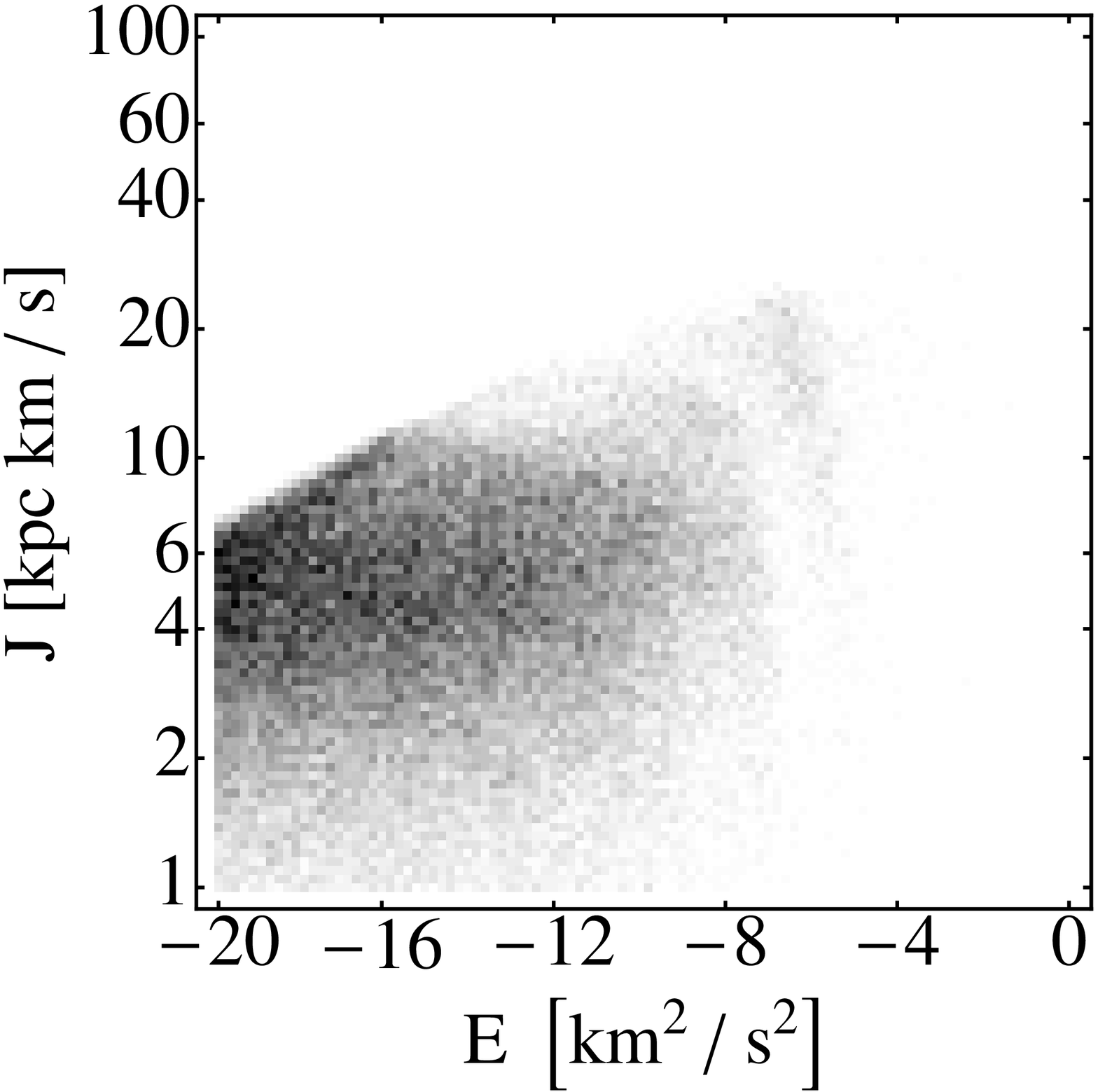}    \\
\end{tabular}
\end{center}
    \caption{The distribution of bound stellar particles in the
	energy-angular momentum plane (darker shading corresponds to the higher density of particles)
        at 6 (left column panels), 8 (middle column) and
	10 (right column) Gyr from the beginning of the simulation. The upper diagrams show
	the results for
	all bound stellar particles, the lower ones only for the stars in the bar. No diagram is
	plotted in the lower right corner because the bar is destroyed by this stage.}
    \label{je}
\end{figure*}

Combining the results shown in the two upper panels of Fig.~\ref{veldisp}
we can study the relation between the stellar velocity dispersion measured at $r_{\rm max}$ and
the maximum circular velocity $V_{\rm max}$ of the simulated dwarf. This relation is used to
construct cumulative velocity functions of dwarf galaxies in the Local Group. Comparisons with the
corresponding functions of dark matter satellites in galaxy-sized dark matter haloes gave rise to
the missing satellites problem. However, the velocity functions of satellites depend sensitively on
the particular assumptions made to infer $V_{\rm max}$ from observations. For example, Moore et al.
(1999) adopted an isothermal halo model with a flat circular velocity profile, while Kravtsov et
al. (2004a) used circular velocities measured either from the rotation curve or from the
line-of-sight velocity dispersion, assuming isotropic velocities.
All these simulations lacked the baryonic component, while here we can assess directly
the relation between $V_{\rm max}$ and the stellar velocity dispersion.
The ratio of the total velocity dispersion
$\sigma_{\rm total} = (\sigma_R^2 + \sigma_z^2 + \sigma_\phi^2)^{1/2}$ and $V_{\rm max}$
is plotted with a solid line in the lower panel of Fig.~\ref{veldisp}. Since the rotation is
not negligible at some stages of the evolution, with dotted line we also plotted the quantity
$\sigma'_{\rm total} = (\sigma^2_{\rm total} + V_\phi^2)^{1/2}$. It turns out that
the measured velocity dispersion traces very well $V_{\rm max}$
for most of the time with $\sigma'_{\rm total}/V_{\rm max} \approx 1$.

Fig.~\ref{shape} shows the evolution of the shape of the stellar component of
the dwarf as a function of time. The two lines plot the ratios of the
principal axis lengths (with $a,b,c$ being the longest, middle and shortest axis
respectively) computed from the eigenvalues of the moment of the inertia tensor using the method
described by Kazantzidis et al. (2004c) again at the radius $r_{\rm max}$ of the maximum circular
velocity (see Fig.~\ref{veldisp}). We see that after the first pericentre the axis ratio does not
significantly change, i.e. the disk is preserved. After the second pericentre a tidally induced
bar is formed in spite of the fact that our initial disk was bar-stable in isolation. The bar
becomes thicker with time with both axis ratios systematically increasing and is finally dissolved
after the last pericentre to form a triaxial galaxy.

\begin{figure*}
    \leavevmode
    \epsfxsize=14cm
    \epsfbox[20 0 544 190]{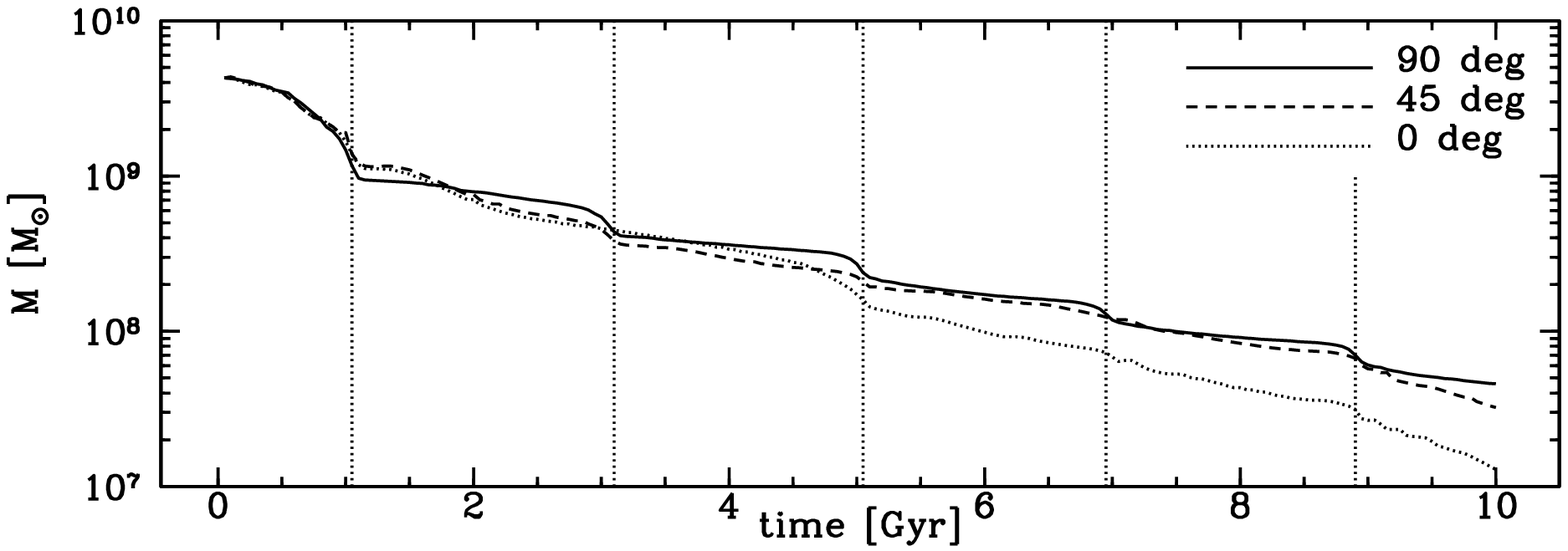}
    \epsfxsize=14cm
    \epsfbox[22 0 518 190]{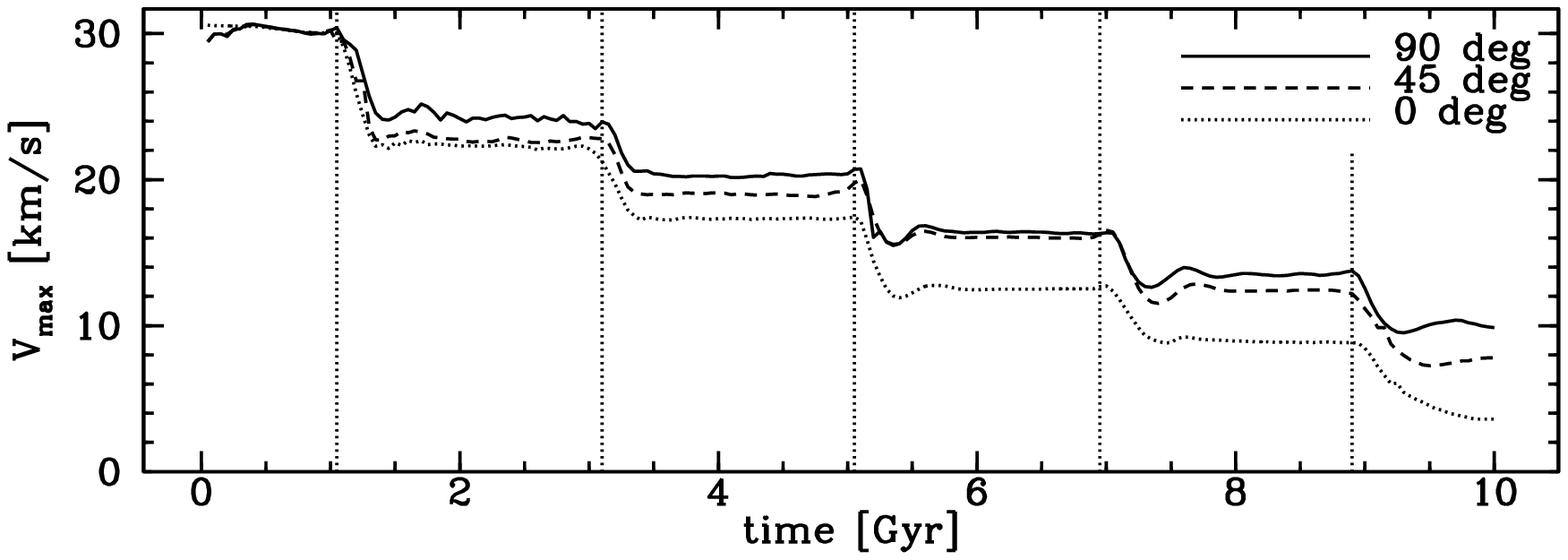}
    \epsfxsize=14cm
    \epsfbox[22 -10 518 180]{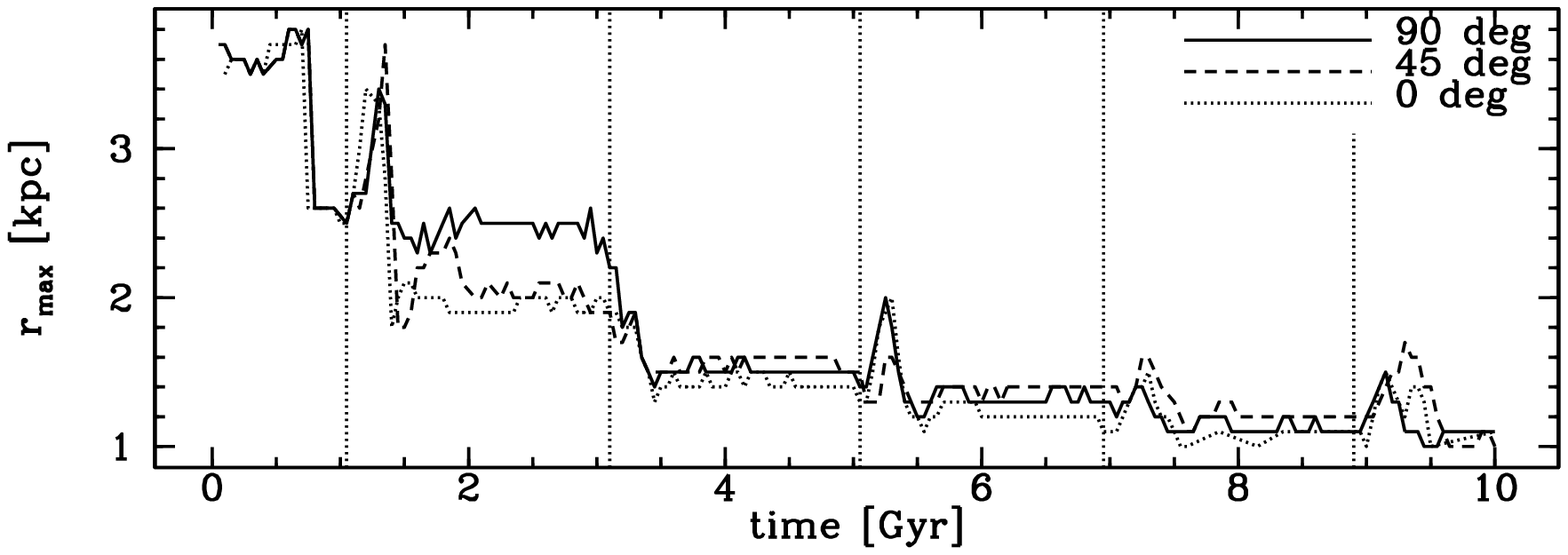}
    \caption{The total bound mass (upper panel) and the maximum circular velocity (middle panel) evolution for
	three different initial inclinations of the disk. The lower panel shows the radius $r_{\rm max}$
	at which the maximum circular velocity occurs.
	In each panel the solid line corresponds to our default inclination of 90 deg while
	the dashed and dotted lines show the results for 45 deg and 0 deg inclination respectively.}
    \label{masses_combined}
\end{figure*}

\begin{figure*}
    \leavevmode
    \epsfxsize=16cm
    \epsfbox{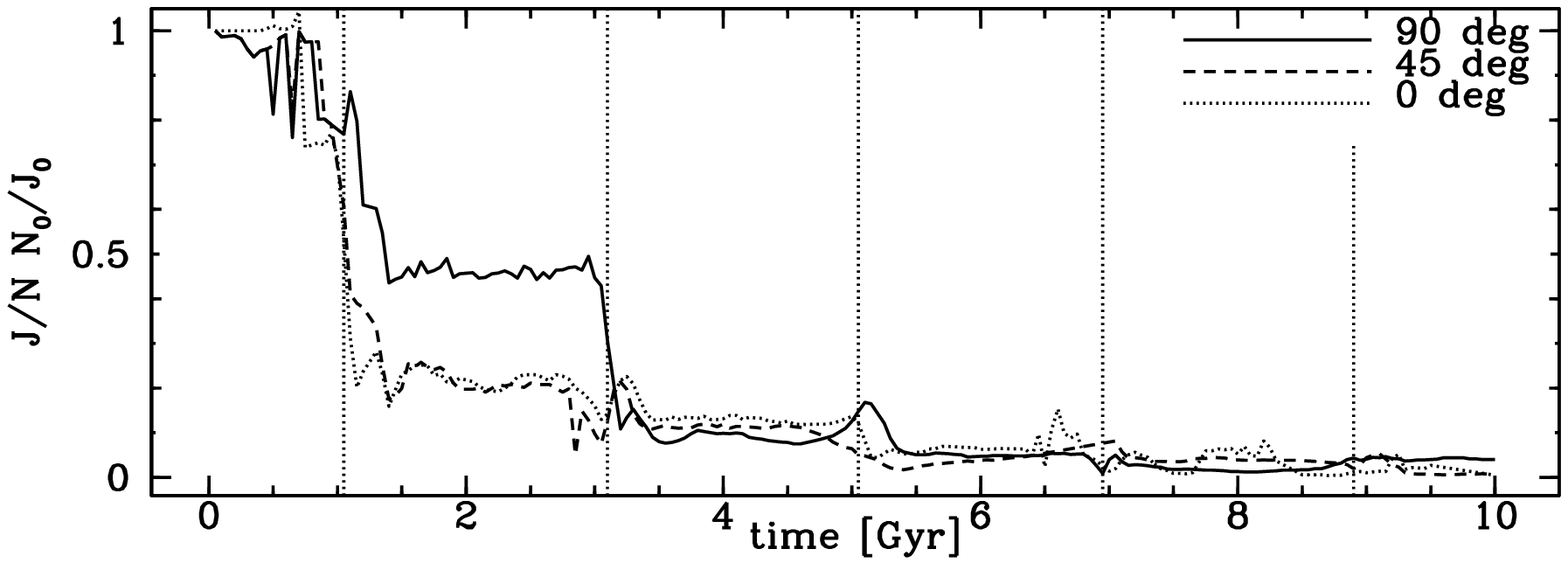}
    \epsfxsize=16cm
    \epsfbox{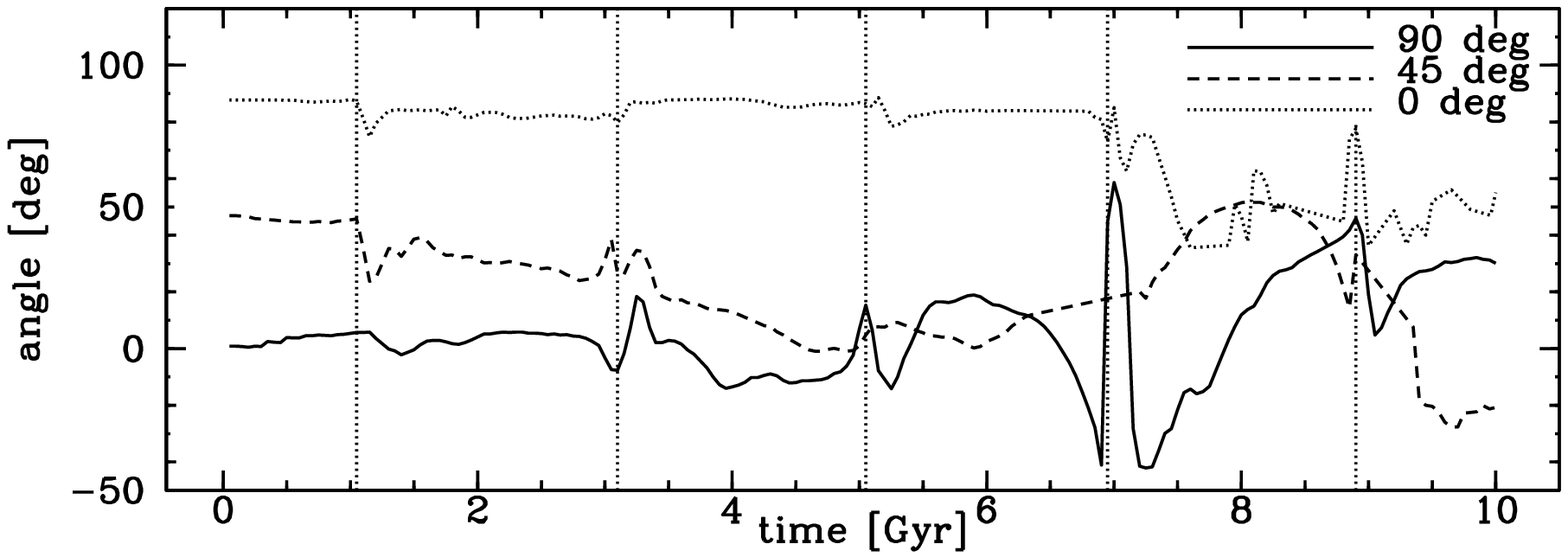}
    \caption{The upper panel shows the scaled mean angular momentum of the stellar particles
	for the three simulations. The lower panel plots the angular deviation of the
	angular momentum vector from the orbital plane.
	In both panels vertical dotted lines indicate pericentre passages.}
    \label{angmom}
\end{figure*}

\begin{figure*}
    \leavevmode
    \epsfxsize=14cm
    \epsfbox[20 0 518 180]{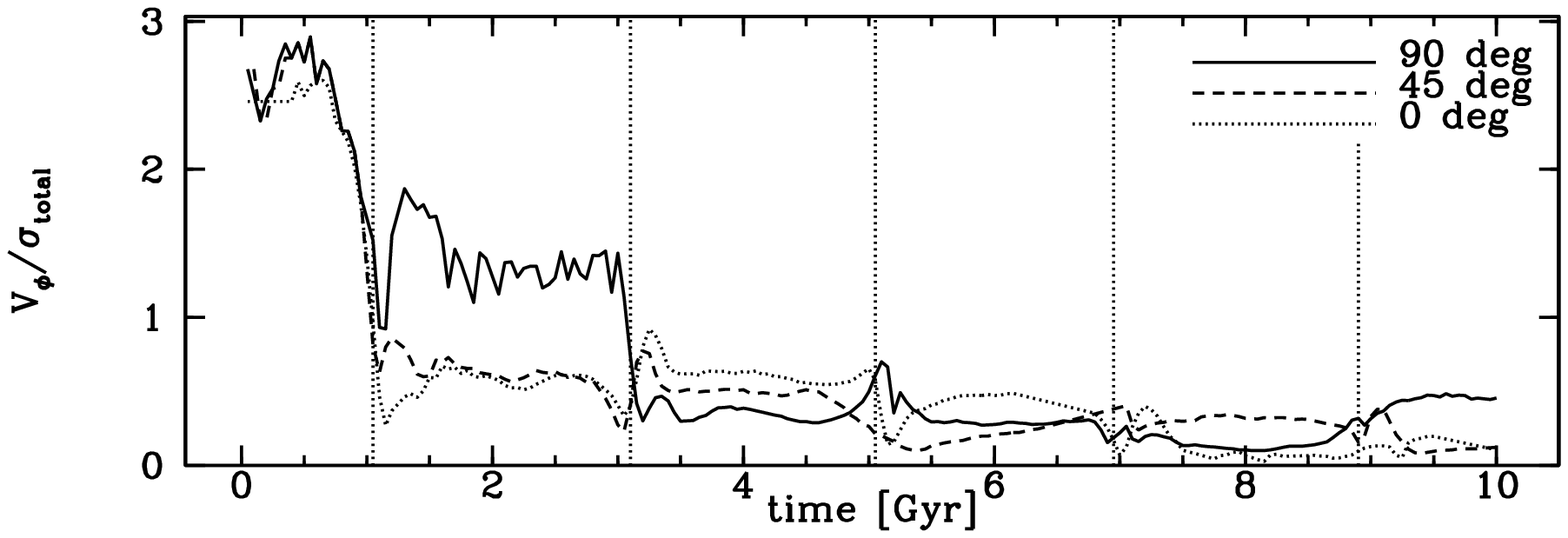}
    \epsfxsize=14cm
    \epsfbox[20 0 518 180]{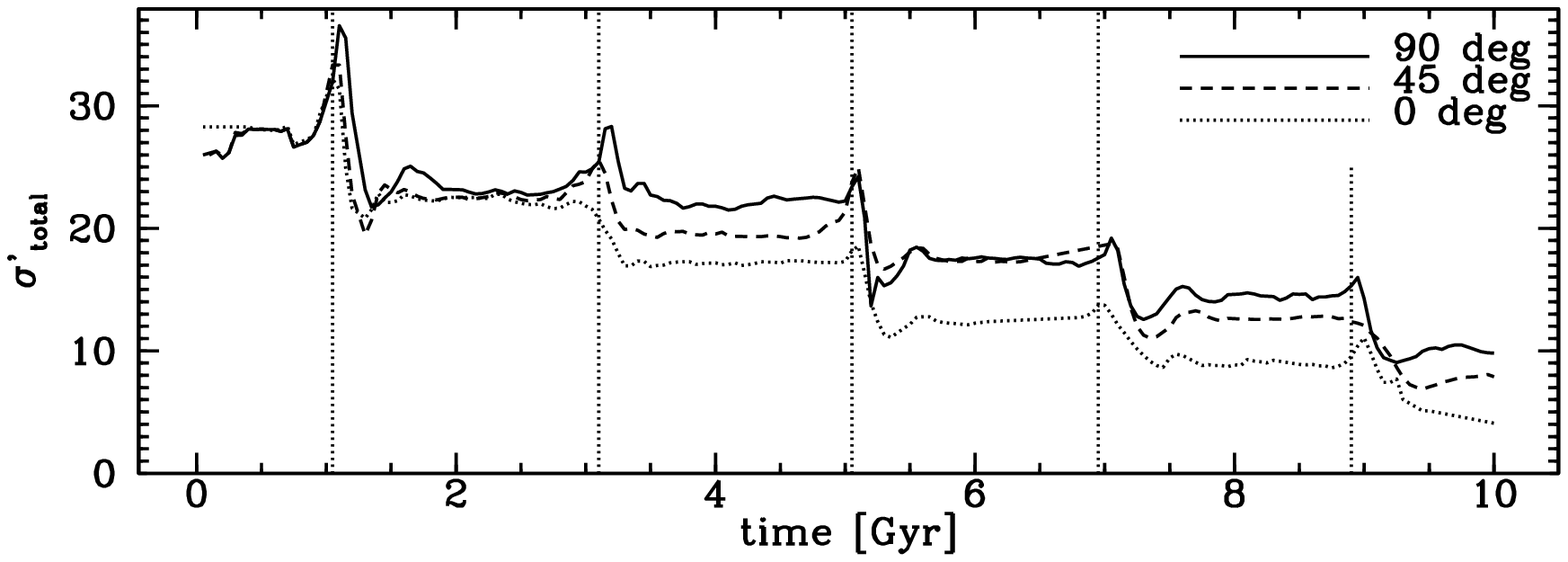}
    \epsfxsize=14cm
    \epsfbox[20 0 518 180]{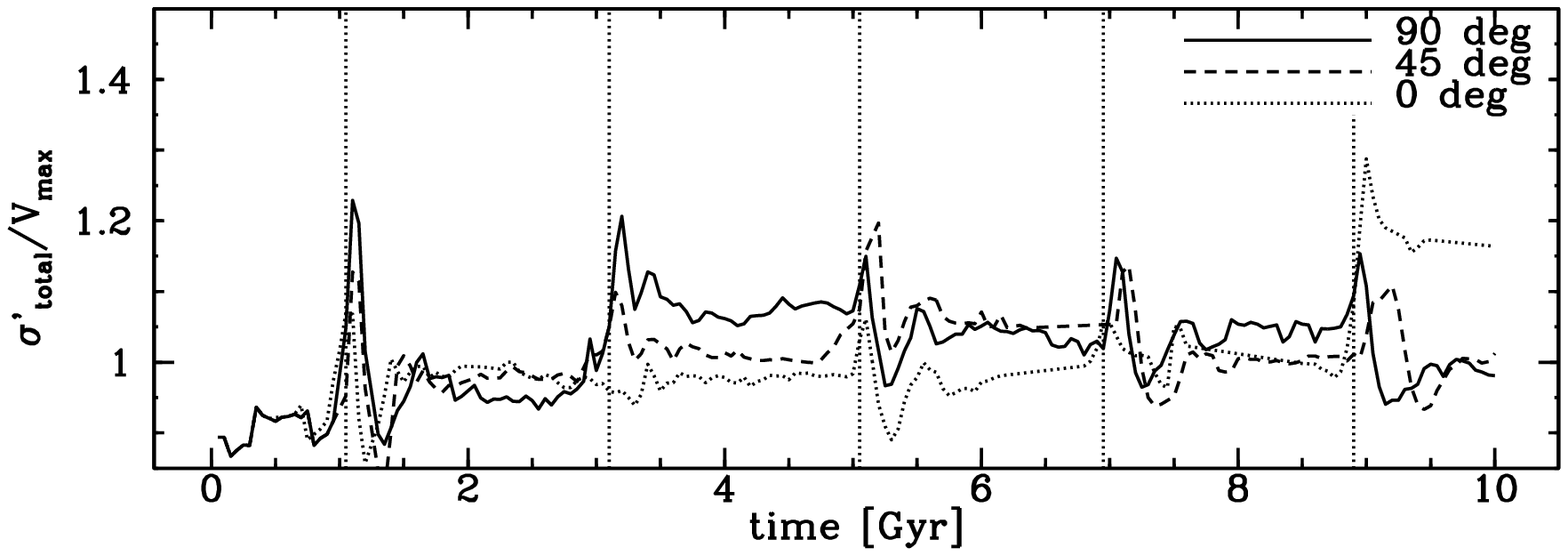}
    \caption{The upper panel shows the evolution of the ratio of the mean rotation velocity
	to the total velocity dispersion (from random motion only) for the three simulations.
	The middle panel plots the evolution of the total velocity dispersion (including the rotation velocity)
	for the three cases. In the lower panel the same quantity is shown in units of $V_{\rm max}$.
	In all panels vertical dotted lines indicate pericentre passages.}
    \label{sigmas}
\end{figure*}

\begin{figure*}
    \leavevmode
    \epsfxsize=16cm
    \epsfbox{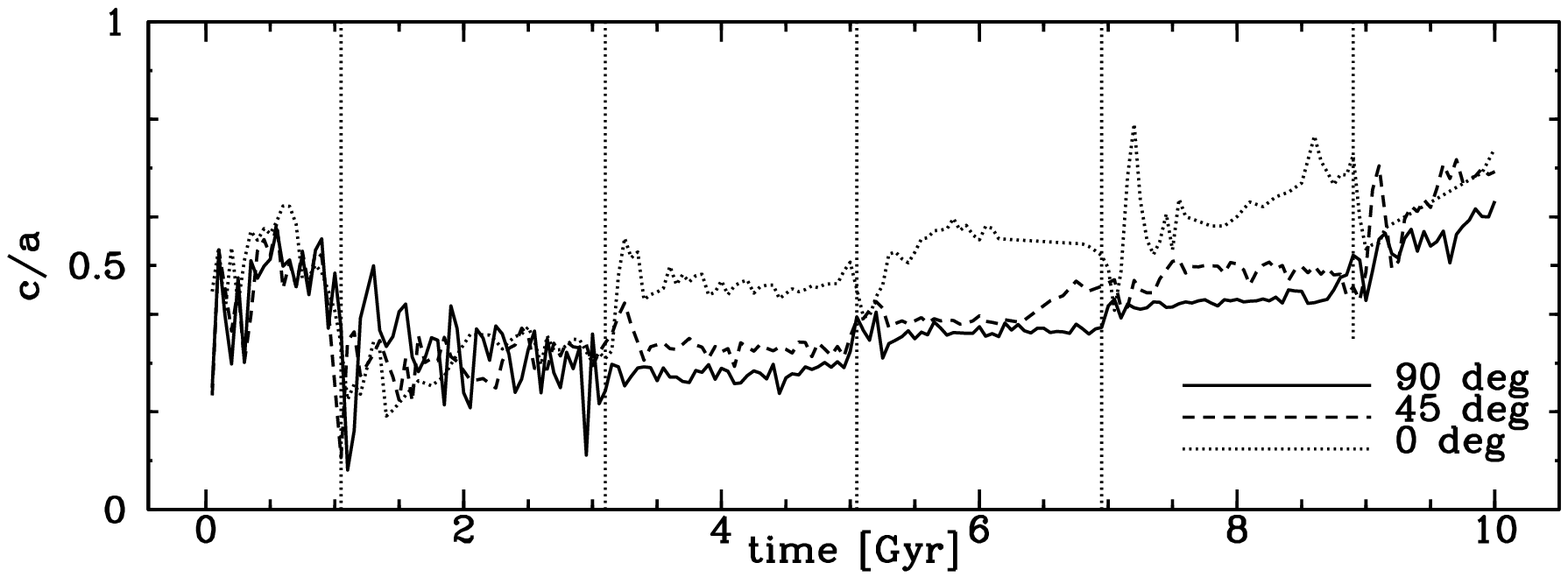}
    \epsfxsize=16cm
    \epsfbox{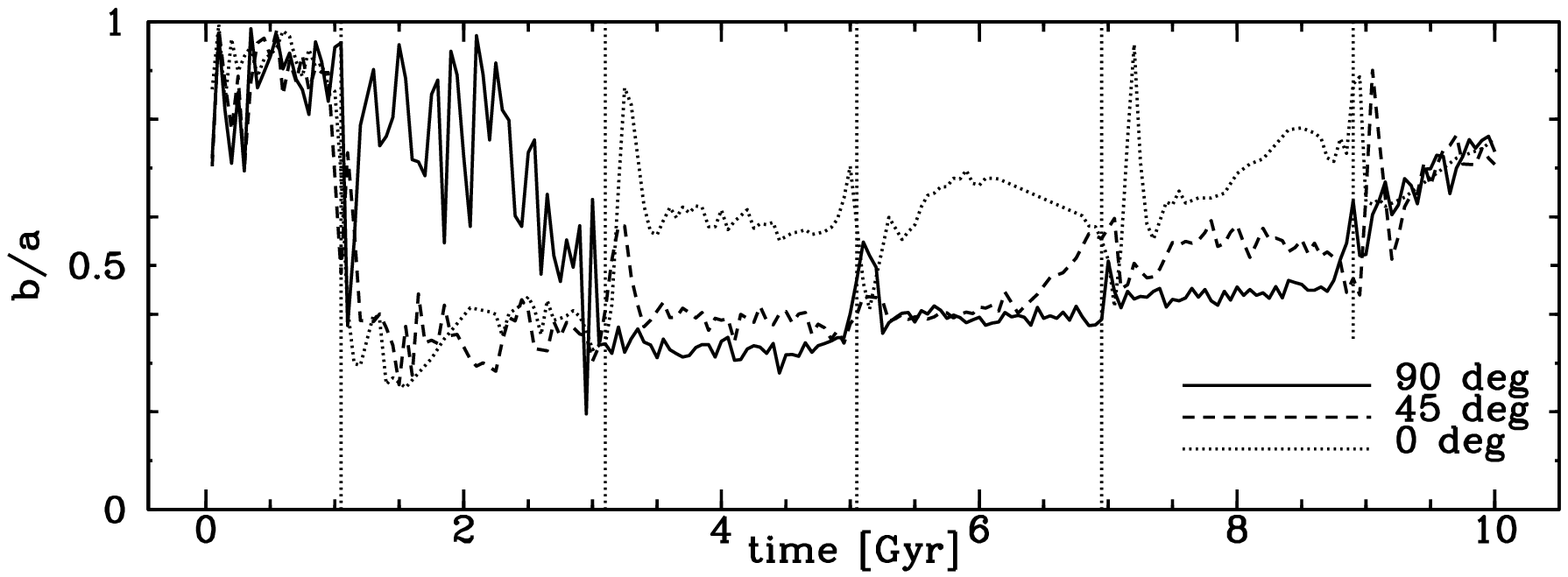}
    \epsfxsize=16cm
    \epsfbox{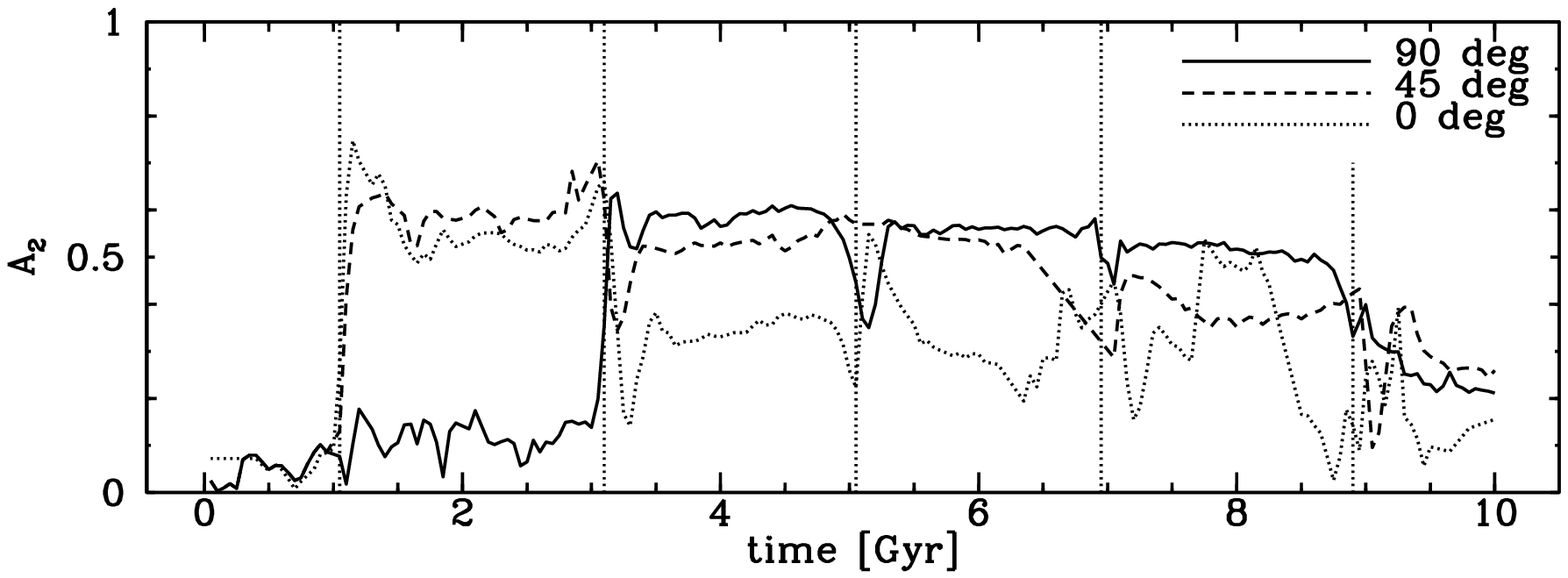}
    \caption{Comparison of the evolution of the shapes of the dwarfs measured at $r_{\rm max}$ corresponding to
	$V_{\rm max}$ in the three simulations. The upper panel shows the ratio of the shortest to the longest axis
	($c/a$), the middle panel the ratio of the intermediate axis to the longest one ($b/a$). In the lower panel
	we plot the amplitude of the bar mode $A_2$.}
    \label{shapes_all}
\end{figure*}

Since the axis ratio alone cannot distinguish between bars and prolate spheroids, the shape of the
dwarf is further illustrated in Fig.~\ref{surface} where we show snapshots from our default
simulation corresponding to the last three apocentres at 6, 8 and 10 Gyr from the beginning of the
simulation (from the left to the right column). In each case we selected stars within $r=2.5$ kpc
from the centre of the dwarf. The upper panels plot the distribution of the stars, while the lower
ones present the contours of the surface density distribution. In each case the observation is made
along the shortest axis of the stellar distribution (in the two earlier snapshots where the two
shorter axes are not distinguishable we chose the one more aligned with the total angular momentum
of the stars). It is clear that except for the latest output the surface density contours depart
strongly from elliptical and are bar-like.

Another feature that distinguishes bars from prolate spheroids is the orbital structure, i.e. a
significant fraction of the orbits in a bar must be radial. A detailed study of this subject is
beyond the scope of the present paper but in Fig.~\ref{je} we show that this is indeed the case.
The Figure presents the distribution of bound stellar particles in the energy-angular momentum plane
again at 6, 8 and 10 Gyr from the beginning of the simulation (from the left to the right column).
In the upper row we show results for all bound stars, in the lower one for stars in the bar. In the
two earlier stages most of the stars occupy the region of low angular momentum corresponding to
radial orbits.

\section{Dependence on the initial orientation of the disk}

In the previous section we presented the analysis of our default simulation, where the stellar disk of the dwarf
is initially positioned perpendicular to the orbital plane (inclination 90 deg).
In order to make our conclusions more general
we reran the simulation with different initial conditions: with the disk of the dwarf inclined by 45 deg
to the orbital plane and lying exactly in the orbital plane (inclination 0 deg).
The orbit and all the other simulation parameters
remained the same. Below we will refer to the three cases respectively as 90, 45 and 0 deg inclination.

The upper panel of Fig.~\ref{masses_combined} shows the evolution of the total bound mass
in the three cases. In all cases the
evolution looks similar in the earlier stages. After about 5 Gyr the mass loss of the 0 deg case speeds up
and the final product is much less massive than in the other two cases. This
behaviour can be explained by resonances between the inner and orbital motion of the stars:
it is easier to eject stars which have angular momenta similar to the
orbital angular momentum of the dwarf, i.e are on more prolate orbits (Read et al. 2006).
Interestingly, the mass-to-light ratios decrease in time in a similar way for
all cases. At the end of the evolution even though the masses differ the mass-to-light ratios are
all of the order of 10 solar units (assuming the stellar mass-to-light ratio of 3 as before).

The middle panel of Fig.~\ref{masses_combined} compares the evolution of the maximum circular velocity $V_{\rm max}$
in time for the three initial disk inclinations. The evolution in all three cases is similar
except for the fact that decreasing the initial inclination angle increases the mass loss. Still, the maximum
circular velocity remains constant between pericentre passages. The lower panel of Fig.~\ref{masses_combined}
plots the values of the radius $r_{\rm max}$ at which $V_{\rm max}$ occurs. Note that the radii are similar for
most of the time, even at later stages where the masses start to differ.

Dwarf spheroidals are low-angular momentum systems, as highlighted by their
low ratio of the rotation velocity to the velocity dispersion, $V_{\phi}/\sigma$ (Mateo 1998).
It is thus important to study the angular momentum evolution of a tidally stirred disky dwarf.
The upper panel of Fig.~\ref{angmom}
compares the evolution of the mean angular momentum of the stellar particles within $r_{\rm max}$. The angular
momenta were expressed in units of the initial values. The results clearly show that the angular momentum per star
systematically decreases in time. We have also noticed that there is no angular momentum transfer in the inner
parts of the dwarf between the stellar component and the dark matter halo. Both show similar behaviour except early
on when the outer parts of the bound dark matter halo gain angular momentum from the stellar particles, but it is
quickly transferred outside and then lost as the outer halo gets stripped. Therefore bound dark matter particles do
not accumulate angular momentum contrary to halo particles in barred spiral galaxies (Athanassoula 2002).
Angular momentum is instead transferred directly outside to the tidal streams.

The lower panel of the Figure shows the angle of deviation of the rotation plane (the plane
perpendicular to the angular momentum vector) from the original inclination of the disk. We immediately
notice that the case of 90 deg is the most unstable in the sense that its rotation plane is most
strongly perturbed. This is understandable since in this case the tidal force acts in the direction
perpendicular to the initial disk. The variation is weaker for other cases, especially for the 0 deg
inclination where the rotation plane remains unchanged almost until the end.

The upper panel of Fig.~\ref{sigmas} presents the ratio of the mean rotation velocity
of stars within $r_{\rm max}$ to the total stellar velocity dispersion, defined as before.
We see that the ratio decreases systematically for all cases. However, the removal of rotation seems to be the
least effective in the 90 deg case, where significant amount of rotation remains even at the end.
The middle panel of Fig.~\ref{sigmas} plots the evolution of the total velocity dispersion
(including the rotation velocity) for the three cases. Since the dispersion is a measure of the mass, its evolution
is similar to the evolution of the total bound mass in Fig.~\ref{masses_combined}. In the lower panel of
Fig.~\ref{sigmas} we plot the total velocity dispersion divided by $V_{\rm max}$. This ratio is constant
throughout the evolution and remains at the level of unity everywhere and for all cases except close to pericentres.
One more important exception appears during the last 1 Gyr of evolution for the case of 0 deg inclination:
the ratio is not constant anymore and its value increases up to about 1.2. In addition, the structure of the galaxy
is very irregular at these stages, clearly signifying a state close to disruption. The structure and kinematics of
the final stages of the galaxies produced in these simulations will be studied in more detail elsewhere ({\L}okas
et al., in preparation).

In Fig.~\ref{shapes_all} we compare the evolution of the shape of the stellar component in the three cases. The
upper panel of the Figure shows the ratio of the shortest to longest axis ($c/a$) and the middle panel plots the
ratio of the middle to the longest axis ($b/a$). The measurements were done as before for stars within the radius
$r_{\rm max}$ at which $V_{\rm max}$ occurs. The main difference is that in the 45 and 0 deg cases the disk is
destroyed and transformed into a bar already after the first pericentre passage while for 90 deg it survived
until the second pericentre. In addition, the axis ratios are significantly larger for 0 deg than
for other cases (the stellar component is more spherical). At the end all dwarfs have spheroidal
shapes with $b/a$ = 0.75 and $c/a=0.6-0.7$. The 90 deg case is more significantly triaxial, while
the 45 and 0 deg cases are truly spheroidal with the two shorter axes almost equal. This picture is
confirmed by the behaviour of the bar amplitude of the Fourier decomposition of stellar phases,
$A_2 = |\Sigma_{j} \exp(2 i \phi_j)|/N$ where $\phi_j$ are the phases of the positions of stellar
particles projected onto the rotation plane. As demonstrated in the lower panel of
Fig.~\ref{shapes_all}, the bar appears to be weaker in the 0 deg case during the evolution, but at
the end it seems similarly diminished in all cases.

The difference in the evolution of the shape after the first pericentre confirms the picture suggested by the
evolution of the angular momentum and rotation velocity: the 90 deg case retained much more rotation than the
other cases. After the second pericentre passage all cases possess a bar, a significant amount of the initial
angular momentum is lost at this stage and the radial velocity dispersion $\sigma_R$ rises dramatically since radial
orbits dominate the elongated bar (see the middle panel of Fig.~\ref{veldisp}). During the following pericentre passages
the angular momentum is transferred outside the galaxy, slowing down the rotation.
The three cases also differ somewhat in the fifth pericentre. After the fourth pericentre passage almost all angular
momentum is lost and there is little rotation. At the fifth pericentre
the 90 deg case gains some angular momentum, while for the other cases it remains on a similar
level (0 deg) or is lost (45 deg). In the simulation started at 90 deg a triaxial galaxy with
significant rotation is formed while in the remaining cases the shape is a prolate spheroid ($b=c=0.7 a$)
and the rotation level is very low.

According to the picture presented here, the rotating bar seems to be stable as
long as angular momentum is transferred outside. After almost
all angular momentum is removed, a spheroidal galaxy is created.
Depending on the efficiency of this process, a spheroidal
galaxy may require more or less time to form or smaller/larger number
of pericentres. The transformation proceeds rather smoothly by systematic shortening of the bar. In the case of 0
deg inclination the transformation seems to be the most efficient and the galaxy can be considered a spheroidal
already after the second pericentre where $b/a=0.6$ and $c/a=0.5$. Our results also suggest that dwarf spheroidal
galaxies could possess some rotation which they had retained from their initial state or even gained in previous
pericentre passages. In this conclusion we agree with Mayer et al. (2001) who showed
that a range of final $V_{\phi}/\sigma$ and structural properties can be obtained by varying the initial dwarf
model and the orbit.

We also note that no bar buckling occurs in this simulation. Bar buckling
which isotropizes the velocity ellipsoid and puffs up the stellar bar was
identified as the main mechanism responsible for the formation of a
dSph in the case of high surface brightness dwarfs (Mayer et al. 2001).
A similar result was found for the transformation of relatively massive
and bright dwarfs in galaxy clusters (Mastropietro et al. 2005).
Therefore, it would seem that there are at least two channels of dSph formation,
tidal heating and shocking of the bar and bar buckling. The first scenario applies to
low surface brightness dwarfs as the ones studied in this work. This mechanism likely is
the dominant one for producing faint dSphs such as those of the Local Group (see also Mayer
et al. 2007). On the other hand, buckling requires a strong bar and thus a
fairly high surface brightness system.

In summary, the formation of a dSph galaxy out of a disk-like system appears to be a long and complicated process.
After the first or second pericentre passage the disk is transformed into a bar which is then systematically
shortened. The mass and angular momentum loss and the transformation of the shape proceeds the faster, the closer
the initial plane of the dwarf galaxy disk is to the orbital plane.
At the end of the simulations, our dSph galaxies are characterized by the following
properties of the stellar component:
\begin{enumerate}
\item low ratio of the rotation velocity to the total velocity dispersion,
$V_{\phi}/\sigma_{\rm total} \le 0.5$,
\item a spheroidal shape with axis ratios $b/a=0.75$ and $c/a=0.6-0.7$,
\item isotropic velocity distribution, $\sigma_R \approx \sigma_z \approx \sigma_{\phi}$.
\end{enumerate}

\section{Observational consequences of tidal stirring}

In this section we discuss some observational characteristics of the tidal stirring model which may help to
distinguish it from other scenarios, e.g. those in which the spheroidal component forms early. If the overall
picture presented above is correct then some of the less evolved dwarfs in the Local Group may still be in
the bar-like phase. Such elongated shapes are indeed observed at least in a few cases, such as Ursa Minor
(e.g. Irwin \& Hatzidimitriou 1995) and the more recently discovered Hercules dwarf (Coleman et al. 2007).
The number of such detections may be low however for a few reasons. First, the bars are oriented
randomly with respect to the observer located at the MW galaxy so some bar-like objects may appear as only
slightly flattened (which is indeed the case with the average ellipticity of dSph galaxy of about 0.3).
Second, the bars may escape detection due to the smoothing procedure usually applied when measuring the
surface density distribution of stars in these objects. Application of a similar procedure to our
simulated data shows that  when the Gaussian smoothing scale is of the order of 10
percent of the radius of the dwarf or larger, the bar may be very difficult to detect even if the line of sight is
perpendicular to the bar. Third, dSph galaxies are always observed against the foreground of the
Milky Way stars which can be only partially removed so no clear impression of the bar will be seen.

\begin{figure*}
    \leavevmode
    \epsfxsize=17cm
    \epsfbox[0 10 485 170]{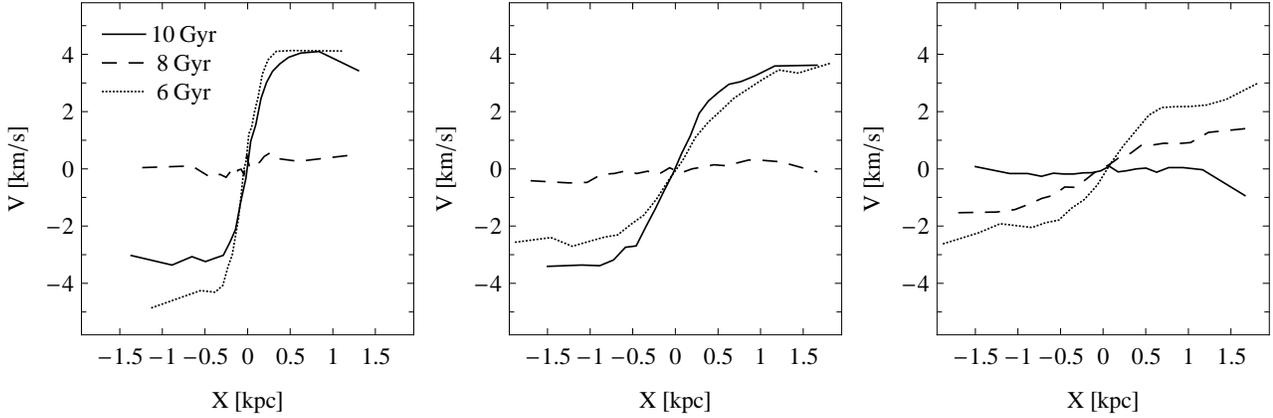}
    \caption{Rotation curves of the dwarf galaxy
	measured along the major axis of the image for the case with 90 deg initial inclination. In each panel the solid,
	dashed and dotted line corresponds to the measurements done at 10, 8 and 6 Gyr from the start of the
	simulation.
	For all outputs the rotation velocities were measured for stars within $r<2.5$ kpc from the
	centre of the dwarf by projecting along the line of sight and binning the data along the major axis taking
	approximately equal number of stars per bin. The three panels from the left to the right correspond
	respectively to observation along the longest, intermediate and the shortest axis of the dwarf.}
    \label{rotation}
\end{figure*}

Another fingerprint of the tidal stirring model is the rotation of the remnants. If the dSph galaxies we observe
today indeed started out as disks they should still, at least in some cases, possess some intrinsic rotation. In
Fig.~\ref{rotation} we show examples of rotation curves measured for our simulated dwarf (again with the 90
deg initial inclination of the disk) at different stages of its evolution: at 6, 8 and 10 Gyr
from the start of the simulation (as in Figs.~\ref{surface} and \ref{je}). For each output we
selected only stars within $r<2.5$ kpc from the centre of the dwarf (to avoid the contamination by
tidal tails) and projected their velocities along the line of sight. In each panel the solid,
dashed and dotted curve shows the rotation curve measured for the dwarf at 10, 8 and 6 Gyr. The
data were then binned into groups containing approximately equal numbers of stars and plotted as a
function of projected distance along the major axis of the projected image. The three panels of
Fig.~\ref{rotation} correspond to observation along the longest, intermediate and the shortest axis
of the dwarf measured in 3D.

The examples shown in Fig.~\ref{rotation} may suggest that intrinsic rotation should be quite easily detected in
majority of configurations. Note however, that we have chosen for the Figure the case with 90 deg inclination
which has significantly more rotation in the final stage (see the upper panel of Fig.~\ref{sigmas})
than the products of simulations with 45 and 0 inclinations, where the rotation would be hardly
detectable (for details see {\L}okas et al., in preparation). Although detection of intrinsic
rotation should provide a conclusive argument for the tidal stirring scenario, such rotation may
also be difficult to distinguish from the velocity gradient induced by the presence of tidal tails
and, for nearby objects, from the velocity gradient caused by transverse motions (so called
`perspective rotation', see Kaplinghat \& Strigari 2008; Walker, Mateo \& Olszewski 2008).
Detection of rotation at different levels of significance has been claimed for a few dSph galaxies,
e.g. for the Ursa Minor (Hargreaves et al. 1994; Armandroff, Olszewski \& Pryor 1995), Sculptor
(Battaglia et al. 2008), Leo I (Sohn et al. 2007; Mateo, Olszewski \& Walker 2008), Cetus (Lewis
et al. 2007) and Tucana (Fraternali et al. 2009). Among these, the case of Leo I seems the most
convincing: due to its large distance the perspective rotation is probably unimportant and the
intrinsic rotation can be distinguished from the one induced by tidal tails because the latter is
in the opposite direction so that the reversal of rotation is observed at larger distances from the
centre of the dwarf ({\L}okas et al. 2008).

\section{Summary and Discussion}

We have studied the tidal evolution of two-component dwarf galaxies on
an eccentric orbit in the MW potential using high-resolution
$N$-body simulations. We have conducted a detailed analysis of how low
surface brightness disk-like systems similar to present-day
dwarf irregulars transform into dSph galaxies. The present investigation
allows for a quantitative understanding of the transformation which improves
considerably on previous work on the subject.
Initially, the tidal forces at pericentre trigger bar formation in the disk, which at the same time
suffers stripping of its outer region. The rotating bar transfers angular momentum outward.
As long as the angular momentum is transferred outside,
the bar-like shape is preserved. The transformation into a dSph occurs by systematic shortening of the bar. The
orbital distribution of the bound stars changes so that nearly radial orbits of the
bar are replaced by more circular orbits. This is different from the bar buckling instability identified in Mayer
et al. (2001) and Mastropietro et al. (2005) as the mechanism whereby high surface brightness, relatively massive
disky dwarfs transform into dSphs.

We find that the velocity dispersions of the stars trace well the maximum
circular velocity $V_{\rm max}$ of the dwarf except close to pericentres. In the spheroidal stage the velocity
distribution is isotropic, but during the whole evolution we find approximately $\sigma'_{\rm total}= V_{\rm
max}$. The 1D velocity dispersion, $\sigma$, corresponding to the line-of-sight velocity dispersion measured in the
observations, is related to $\sigma'_{\rm total}$ so that $\sigma=\sigma'_{\rm total}/\sqrt{3}$ and we find the
approximate relation $\sigma \sim V_{\rm max}/\sqrt{3}$. We also find that $V_{\rm max}$ traces reasonably well the
total bound mass of the dwarf during the entire evolution. Assuming that the line-of-sight velocity dispersion can
be reliably measured from kinematic data, the above facts suggest a new independent way of measuring the masses of
dSphs.

These findings have intriguing implications for the missing satellites problem, which is normally formulated
by measuring the number of subhaloes with a given value of $V_{\rm max}$. In particular, it suggests
that circular velocities of observed dwarfs derived from either a simple
isothermal halo model with a flat circular velocity profile (Moore et
al. 1999) or line-of-sight velocity dispersions, assuming isotropic velocities
(Klypin et al. 1999) were reasonable. This is true at least for dSphs that have
a moderate mass-to-light ratio as our simulated dwarf (e.g. Fornax or Leo I). Dwarfs embedded in
a much more massive dark matter halo may in principle have a different relation between $\sigma$ and $V_{\rm max}$
since the stars could probe a region well inside the radius at which $V_{\rm max}$ occurs. Kazantzidis et
al. (2004b) found that the stellar kinematics of Draco, that has $\sigma \sim 10-12$ km s$^{-1}$, can be reproduced
in haloes with $V_{\rm max} \sim 25-30$ km s$^{-1}$, which would imply $\sigma/V_{\rm max} \sim 0.3-0.5$.
A more recent study of Draco by {\L}okas, Mamon \& Prada (2005),
which took into account the contamination of the sample by tidal tails and included the fourth velocity
moment in the analysis, found $\sigma \sim 8$ km s$^{-1}$ and $V_{\rm max} \sim 16$ km s$^{-1}$ again in
agreement with the relation found here.

We find that the mass-to-light ratio of the dwarfs drops significantly at the beginning of the evolution which is
due to stripping of the extended dark halo. However, the ratio does not change significantly during the middle and
late stages of the evolution and is also independent of the initial inclination of the disk. This suggests that
very high values of dark matter content in some of the dSph galaxies, such as Draco or Ursa Major, are not an
effect of tidal evolution but are rather the result of the formation process of their progenitors, or are caused by
other mechanisms that affected their baryonic mass fraction. Photoevaporation of the gas after reionization (Babul
\& Rees 1992; Bullock, Kravtsov \& Weinberg 2000; Barkana \& Loeb 2001) could not cause the reduced baryon
fraction. In fact our numerical experiment clearly shows that the original halo mass and $V_{\rm max}$ of a dSph
produced by tidal stirring was significantly higher than it is today, and was well above the threshold below which
photoevaporation is effective (this requires  $V_{\rm max} < 15$ km s$^{-1}$ according to the radiative transfer
simulations of Susa \& Umemura 2004). Instead, the combined effect of efficient heating by the cosmic ultraviolet
background as the satellites fell into the MW halo at $z > 1$ and the ram pressure stripping in the gaseous corona
around the MW disk might have removed most of the baryons, which remained in the gas phase rather than turning into
stars (Mayer et al. 2007).

Recently Pe\~narrubia et al. (2008) studied tidal evolution of dSphs in a host
potential assuming a King model for the stellar component. If the picture
presented here is correct, then their assumption might not be very realistic,
as in our case the dSph galaxy forms rather late in the evolution.
In particular, Pe\~narrubia et al. (2008) find that the result of
the tidal evolution is to increase the mass-to-light ratio. This is due to the fact that the
stars in their model are distributed according to the King profile with a core, so that they are
loosely bound and the stellar component is much more heavily stripped than in
our case. In our simulations, the initial conditions and the bar present for most of the time
enhances the resilience of the stellar component to tidal stripping by
increasing the depth of the potential well.

One potential problem which
arises from our study is the fact that not many bar-like structures are observed in the Local Group
except for the Large Magellanic Cloud (LMC), the Ursa Minor dSph and the Hercules dwarf. It is not clear whether
structures of this kind are detectable in such low surface brightness systems, but the simulations predict that
they should be quite common as this is likely to be the longest stage of the dwarf evolution. It is possible that
at least several observed dSph galaxies could in reality be in a bar-like stage with the bar aligned along the
line-of-sight and seen as spheroids due to projection effects. As we have shown, bars may also escape detection due
to smoothing procedures usually applied. The LMC, which is known to have a more extended orbit than in the case of
our simulated dwarf (Kallivayalil, van der Marel \& Alcock 2006), consists of a bar and a thin stellar disk. Also,
orbital evolution models using 3D velocities constrained by recent proper motion measurements suggest that the LMC
may currently be on its first passage about the MW (Besla et al. 2007). The above confirm the picture we have
presented, as we expect LMC to be less evolved and thus in some transitory stage between a stellar disk and a
spheroid. However, most of the MW companions might indeed have had enough time to evolve into spheroidals since
galaxy-sized haloes form early in CDM models and the MW halo was probably already in place at $z=2$ (Governato et
al. 2004, 2007). In addition, the present distances of most dSphs suggest that they were mostly accreted early
(Mayer et al. 2007; Diemand et al. 2007). The situation is different in galaxy clusters which formed recently so
the dwarfs did not have time to complete many orbits. Interestingly, many dwarfs in a transitional stage between a
disky dwarf and a true dSph have been recently discovered in such environments (Lisker et al. 2007).

In this paper we have presented a self-consistent model for the evolution of dwarf galaxies starting from
the most natural progenitor expected in the CDM paradigm.
This consists of a stellar disk embedded in a dark matter halo with
the NFW profile. Cosmological simulations support the idea that isolated dwarfs are
rotating disks, while pressure supported dwarfs are more abundant closer to
the primary galaxy (Kravtsov et al. 2004a; Mayer 2005).
We have shown how the evolution of such objects proceeds and how they are affected by the tidal forces of the host
galaxy. The morphological evolution of dwarf galaxies appears to be a dynamically rich process with several stages
characterized by different events. In this picture recent high resolution simulations where a spheroidal object is
postulated from the beginning (e.g. Pe\~narrubia et al. 2008; Mu\~noz, Majewski \& Johnston 2008) might not be
adequate to describe how dSphs evolved under the action of the tidal field. In fact our results suggest that in
most cases they could have transformed into spheroidal objects quite recently.

The growing kinematic data sets for dSph galaxies should soon make detailed comparisons
between the models and the data possible. For example, our model predicts that if dSphs
formed from disks some residual rotation could still be present in the final stage.
Such rotation has been claimed e.g. for the Ursa Minor dSph galaxy (Hargreaves et al. 1994;
Armandroff, Olszewski \& Pryor 1995).
Recently {\L}okas et al. (2008) have shown that the model presented here
can be used to explain the rotation curve of the Leo I dSph galaxy. It
turns out that in this case not only the internal rotation is present but also the tidal tails
affect the rotation curve in such a way that the rotation is reversed. If
this picture is correct than it provides another strong argument that the results presented
in this paper are relevant to at least some of dSph galaxies.

We argue that earlier attempts to model dSph galaxies as unbound and dark matter-free
stellar remnants (Kroupa 1997) are ruled out based on our results. The tidal
stirring model indeed demonstrates how tidal stripping and the formation of
tidal tails naturally coexist with a substantial bound stellar component
embedded in a relatively massive CDM halo even after several Gyr. Our results
highlight the fact that tidal effects and the existence of a gravitationally
bound dSph galaxy with a relatively high mass-to-light ratio are not mutually
exclusive.

\section*{Acknowledgments}

We wish to thank S. Gottl\"{o}ber and I. Shlosman for discussions.
JK and E{\L} are grateful for the hospitality of
the Institut d'Astrophysique de Paris where part of this work was done.
SK is funded by the Center for
Cosmology and Astro-Particle Physics at The Ohio State University.
The numerical simulations were performed on the
zBox2 supercomputer at the University of Z\"urich.
This research was partially supported by the
Polish Ministry of Science and Higher Education
under grant N N203 0253 33 and the Jumelage program
Astronomie France Pologne of CNRS/PAN.

\end{document}